\documentclass[final,5p,times,twocolumn,authoryear]{elsarticle}

\usepackage[authoryear]{natbib}
\usepackage{graphicx}
\usepackage{subcaption}
\usepackage{amssymb,amsmath}
\usepackage{pdflscape}

\usepackage{lipsum}
\usepackage{epsfig}

\usepackage[utf8]{inputenc}
\usepackage{newunicodechar}
\usepackage{textgreek}

\usepackage{amsthm}

\journal{High Energy Astrophysics}
\UseRawInputEncoding

\begin{document}

\begin{frontmatter}



\title{Probing Broadband Spectral Energy Distribution and Variability of Mrk\,501 in the low flux state}


\author{Javaid Tantry\corref{cor1}\fnref{label1}}
\ead{javaidtantray9@gmail.com}
\affiliation[label1]{Department of Physics, University of Kashmir, Srinagar 190006, India}
\cortext[cor1]{Corresponding author: Javaid Tantry}

\author{Zahir Shah\corref{cor1}\fnref{label2}}
\ead{shahzahir4@gmail.com}
\affiliation[label2]{Department of Physics, Central University of Kashmir, Ganderbal-19113, India}
\cortext[cor1]{Corresponding author: Zahir Shah}

\author{Ranjeev Misra\fnref{label3}}
\affiliation[label3]{Inter-University Centre for Astronomy and Astrophysics,  Post Bag  4, Ganeshkhind, Pune-411007, India}

\author{Naseer Iqbal\fnref{label1}}

\author{Sikandar Akbar\fnref{label1}}

\begin{abstract}
We conducted a multi-wavelength analysis of the blazar Mrk\,501, utilizing observations from \emph{Astro}Sat (SXT, LAXPC), \emph{Swift-UVOT}, and \emph{Fermi-LAT} during the period  August 15, 2016 to March 27, 2022. The resulting multi-wavelength light curve revealed relatively low activity of the source across the electromagnetic spectrum. Notably, logparabola and broken power-law models provided a better fit to the joint X-ray spectra from \emph{Astro}Sat-SXT/LAXPC instruments compared to the power-law model. During the low activity state, the source showed the characteristic “harder when brighter” trend at the X-ray energies. To gain insights into underlying physical processes responsible for the broadband emission, we performed a detailed broadband spectral analysis using the convolved one-zone leptonic model with different forms of particle distributions such as logparabola (LP), broken power-law (BPL), power-law model with maximum energy ($\xi_{max}$), and energy-dependent acceleration (EDA) models. Our analysis revealed similar reduced-$\chi^2$ values for the four particle distributions. The LP and EDA models exhibited the lowest jet powers.
The correlation analyses conducted for the LP and BPL models revealed that there is a positive correlation between jet power and bulk Lorentz factor.
Specifically, in the LP model, jet power proved independent of $\gamma_{min}$, whereas in the broken power-law model, jet power decreased with an increase in $\gamma_{min}$. The jet power in the LP/EDA particle distribution is nearly 10 percent of the Eddington luminosity of a $10^7$ M$_\odot$ black hole. This result suggests that the jet could potentially be fueled by accretion processes.
\end{abstract}


\begin{keyword}
galaxies: active \sep galaxies:  BL Lacertae objects: Mrk 501 \sep jets \sep radiation mechanisms: non-thermal - gamma-rays \sep galaxies:  Jets; Active


\end{keyword}

\end{frontmatter}




\section{Introduction}
\label{introduction}
Blazars are subclass of active galactic nuclei (AGN) with relativistic jet pointing close to the line of sight of an observer \citep{blandford1978, 1995PASP..107..803U}. They are characterized by strong $\gamma-$ray emissions, high degree polarization, rapid variability and nonthermal spectrum extending from radio to $\gamma$-ray energies \citep{wills1992, ackermann2015, Fan2018, 2020ApJ...892..105A, Zhang_2021}. The close pointing of relativistic jet results in relativistic beaming which further amplifies the observational properties \citep{blandford1978,1995PASP..107..803U, 2003ApJ...596..847B}. On the basis of their spectral features, blazars are broadly identified as flat-spectrum radio quasars (FSRQs) and BL Lacertae objects (BL Lacs). FSRQs exhibit strong emission line features in their optical spectrum, while BL\,Lacs display weak or no emission line features \citep{1995PASP..107..803U}.
 
 The broadband spectral energy distribution (SED) of blazars exhibits double humped shape and the emission extends from radio to GeV/TeV energies. According to the leptonic model, this broadband emission can be attributed to the synchrotron 
 and inverse-Compton (IC) processes. The low-energy hump, peaking in the Optical/UV/X-ray energy is produced by the synchrotron process, while
the high energy hump peaking in the $\gamma$-ray band, is mainly explained by IC scattering of low-energy photons \citep{1982ApJ...253...38U, 1995ApJ...441...79B}. The seed photons for IC scattering can be
either synchrotron photons leading to the synchrotron self-Compton process \citep[SSC;][]{1974ApJ...188..353J, Shah_2017}, or photons entering external to the jet called external Compton process 
 \citep[EC;][]{1992A&A...256L..27D, 1994ApJ...421..153S}. Alternatively, the high energy emission from blazars can also be explained through a hadronic process initiated by the relativistic protons, involving the proton-synchrotron processes
and pion production processes \citep{1992A&A...253L..21M, 2001APh....15..121M}.
Based on the peak frequency of the synchrotron component, BL Lacs are further divided into low-energy peaked BL Lac (LBL;  $\nu_p <$  $10^{14}$ Hz), intermediate-energy peaked BL lacs (IBL; $10^{14} <$ $\nu_p <$  $10^{15}$) and high-energy peaked BL Lacs (HBL; $\nu_p >$  $10^{15}$) \citep{1995ApJ...444..567P}.\\ 
 Mrk\,501  is a BL\,Lac object at a redshift $z\sim0.034$. It is one of the nearest bright  HBL source in X-rays \citep{2004A&A...422..103M}. The source was detected at the very high energy (VHE) band by the Whipple observatory in 1995 and became the second extragalactic source detected in the VHE band  \citep{1996ApJ...456L..83Q}.  
The source underwent a strong VHE flare in 1997 with peak flux upto ten times the Crab nebula flux and flux-doubling timescales of half a day \citep{1999A&A...342...69A}. The VHE flare was accompanied by the correlated emissions in the X-ray band with a very hard X-ray spectrum \citep{1998ApJ...492L..17P}.
It showed TeV flare again in June 1998, during which the peak of the synchrotron component shifted to high energy $\geq 50$\,keV \citep{Sambruna2000}. \citet{2007AIPC..921..153P} showed that the VHE flux of the source varied by an order of magnitude during the MAGIC campaign in May--July 2005, with the flux doubling timescale of $\sim 2$ minutes. 

In the X-ray band, Rossi X-Ray Timing Explorer (RXTE) has observed remarkable X-ray variability from the source \citep{2006ApJ...646...61G}.
Later \emph{Swift}-XRT observations showed a powerful and prolonged X-ray activity from the source during  March--October 2014  \citep{10.1093/mnras/stx891}. This long outburst was reported to be superimposed with several shorter flares. Moreover, during a strong flaring event, the X-ray spectral evolution demonstrated a harder when brighter pattern with a spectral index harder than 1.70.  
The source showed another long flare in 2021, a detailed X-ray timing and spectral analysis of this outburst was carried out by \cite{Kapanadze_2023}. During this period, the  X-ray spectrum was hard and showed significant curvature features.

Being a strong variable source, it has been extensively studied in the last three decades. Several multi-wavelength studies of the source have been carried out to understand the physical processes responsible for the flux variation \citep{1998ApJ...492L..17P, Sambruna2000, Abdo_2011}. Recently, \citet{Abe:2024eba}  carried the first multi-wavelength analysis of Mrk\,501 along with the simultaneous X-ray polarisation data from the Imaging X-ray Polarimetry Explorer (IXPE).  The authors explained the variation in the three IXPE pointing observation due to the change in the size of emission region or variation in the magnetic field. 
 The broadband emission of Mrk\,501 during the low flux state (2017 -- 2020) was studied by \citet{abe2023multi}. During this period, a historical low activity is reported in Mrk\,501 in the X-ray and VHE $\gamma$-ray bands.  Such observations helped to unravel the baseline emission from the source. The authors showed that the multi-wavelength light curves are significantly correlated, supporting the leptonic scenario for the broadband emission. The broadband SED during the high and low flux states are adequately modeled with a one-zone leptonic model involving synchrotron and SSC processes \citep{2018A&A...620A.181A, abe2023multi}. 
Similarly, a one-zone SSC model described the broadband SED of Mrk\,501 during an extreme X-ray outburst in July 2014 \citep{2020A&A...637A..86M}. The SEDs in different flux states were reproduced with the variation in the break energy, magnetic field, and spectral shapes of the particle distribution. Especially, on the day of peak X-ray flux, an unusually narrow feature was observed in the VHE spectrum. This feature, inconsistent with classical particle distribution forms, was best described by a logparabola with an additional narrow component, challenging conventional models at TeV energies \citep{2020A&A...637A..86M}. These results suggest that despite numerous studies on the source, the nature of this object is still not well understood.

Blazars represent extremely powerful sources within the extra-galactic universe. Their emission is mainly dominated by the non-thermal emission from the relativistic jets. Theoretical models suggest that these jets originate from a spinning black hole, with their power linked either to the spin and mass of the black hole \citep{10.1093/mnras/179.3.433} or to the black hole spin and the angular velocity \citep{10.1093/mnras/283.3.854,1997MNRAS.292..887G}. In the case of blazars, the jet power is usually estimated through broadband SED modeling.
Understanding the physical processes behind the extreme emission from the relativistic jet remains the main research area in blazars. Several multi-wavelength (MWL) studies have been carried out to understand the emission process and to infer the physical parameters playing a role in the jet emission \citep{2013ApJ...774...18Z, 2014ApJ...789...66Z, Fan_2018}.   
 For example, \citet{Shah_2017} used the approximate analytical expressions for synchrotron, SSC and EC processes to constraint the high energy emission mechanism and source of seed photons for the IC process. Additionally, \citet{article} examined the physical processes behind the high-energy emission from blazars by investigating the correlations between the jet power and other observed properties such as the $\gamma-$ray brightness and the synchrotron luminosity. 
We conduct a multi-wavelength study of Mrk\,501 by using the broadband observations from \emph{Astro}Sat, \emph{Swift}, and \emph{Fermi} Telescopes. Within the framework of one-zone leptonic model, we examined the implications on the jet energetics by using different forms of particle distribution. Particularly, we constrained the underlying particle distribution by obtaining the minimum jet power condition. The correlation analyses are carried out between the jet power and the underlying physical parameters to check the consistency of the particle distributions. 
Recently, \citet{10.1093/mnras/stae706} studied the jet power of Mrk\,501 during its low state, using multi-wavelength data that was not strictly simultaneous, with typical differences of a few hours between \emph{Swift} and \emph{NuSTAR} observations. In contrast, our work utilized the simultaneous multi-wavelength capabilities of \emph{Astro}Sat's SXT and LAXPC instruments in the broad range of X-rays, providing comprehensive coverage of the synchrotron component within the broadband SED \citep{2017A&A...603A..31A}. The high exposure time with \emph{Astro}Sat provides a significantly better quality spectrum, enabling verification of \cite{10.1093/mnras/stae706} results using independent data. Furthermore, longer \emph{Astro}Sat observations allow us to do a detailed temporal and spectral analysis in X-rays, which was not performed in \citet{10.1093/mnras/stae706}.
This work is structured in the following ways: Section \S\ref{sec:style} provides the details of observation and data processing techniques. In section \S\ref{astr:res}, we report the temporal and spectral results of Mrk\,501 using the {\emph Astro}Sat observations. Section \S\ref{mul:anl} contains the multi-wavelength light curve and fractional variability results. The broadband spectral analysis results are presented in Section \S\ref{broad:band}. Finally, the paper concludes with a summary and discussion in Section \S \ref{sum:dis}.
\section{Observations and Data reduction} \label{sec:style}
\subsection {\emph{Astro}Sat}
\emph{Astro}Sat being a multi-wavelength observatory consisting of Ultra-violet
Imaging Telescope \citep[UVIT;][]{2017JApA...38...28T}, Soft X-ray
Focusing Telescope \citep[SXT;][]{2017JApA...38...29S}, Large Area
X-ray Proportional Counter \citep[LAXPC;][]{2016SPIE.9905E..1DY}, and
Cadmium Zinc Telluride Imager \citep[CZTI;][]{2017CSci..113..595R}. It  permits  the
simultaneous observation of the source at UV, soft X-ray, and hard
X-ray energies.
\emph{Astro}Sat conducted five observations of Mrk\,501 during the period MJD\,57615.2 -- 59666, we identified these observations as  S1 (57615.2 -- 57615.9), S2 (58594.9 -- 58596.2), S3 (58910.3 -- 58912), S4 (58934.5 -- 58939.8), and S5 (59662.3 -- 59665.8) . The details of \emph{Astro}Sat observations of Mrk\,501 are given in Table~\ref{tab:1}. In this section, we provide the details of the data reduction procedures applied to the observations made by the SXT and LAXPC instruments.\\
\subsubsection{SXT DATA} 
SXT onboard \emph{Astro}Sat is an X-ray imaging telescope sensitive in the energy range 0.3 -- 8.0\,keV with an effective area of $\sim \rm 90 \,cm^2$ at 1.5\,keV  \citep{2017JApA...38...29S}. It is designed to provide soft X-ray images, light curves and spectra in the energy range 0.3 -- 8.0\,keV. During the time period  MJD 57615 -- 59666, Mrk\,501 was observed in photon counting (PC) mode. We
processed the level-1 data in each orbit to level-2 data by using SXTPIPELINE version AS1SXTLevel2-1.4a released on January 6, 2017. 
In order to obtain a single cleaned event file, we merge the level-2 data of all orbits by using the SXTEVTMERGER tool. Science products (light curve and spectra) were generated from the merged event files
using XSELECT (V2.4d) package which is available in the HEASoft software. A circular region of $15'$ radius centered on the source was used to extract source spectra and light curve, the selected circular region covers  90$\%$ of the source photons. The spectral analysis was carried out
by using the background spectrum “$\rm SkyBkg_-comb_-EL3p5_-Cl_-Rd16p0 v01.pha$”,  and  response matrix file (RMF) $\rm ``sxt _-pc
mat_-g0to12.rmf$" provided by the SXT POC team. Further, we used an off-axis auxiliary response file (ARF) generated using SXT ARF generation tool\footnote{$\rm http://www.tifr.res.in/^\sim astrosat_-sxt/dataanalysis.html$}. To ensure the optimal number of counts per bin, we have used \textit{“ftgrouppha”} package available in HeaSoft software \citep{Kaastra2016}.\\
\subsubsection{LAXPC DATA}
The LAXPC is another payload onboard \emph{Astro}Sat, it consists of 
three co-aligned identical non-imaging proportional counters namely LAXPC10, LAXPC20, and LAXPC30 \citep{2016SPIE.9905E..1DY}. It has a total effective area of 6000 $cm^2$ at 15\,keV. The LAXPC detectors can observe photons with a time resolution of $\sim$ 10 $\mu$s. LAXPC has the ability to perform observations in a broad range of energies 3.0 -- 80.0\,keV. The detailed information about the LAXPC instrument is provided as  \citep{Roy_2016,2016SPIE.9905E..1DY}, and \citep{2017JApA...38...30A}. Due to gain instability issues with  LAXPC10 and LAXPC30 being officially shut down, we have only used data from LAXPC20 in our work. The processing of Level-1 data and the extraction of light curves/spectra were done with the software package LAXPCSOFT. Since blazars are faint sources for LAXPC, we used the scheme of
faint source algorithm which is implemented as a part of LAXPCSOFT for extracting
the light curves and spectra. The background subtraction was done by the standard FTool lcmath package. We restricted the temporal analysis to the energy range 3.0 -- 30.0\,keV and spectral analysis to the energy range 4.0 -- 17.0\,keV, since the background dominates the source spectrum above these energies.
\\

\subsection{\emph{Swift}-UVOT}
\emph{Swift} is a space-based observatory equipped with Burst Alert Telescope (BAT), X-ray Telescope (XRT) \citep{2005SSRv..120..143B,2005SSRv..120..165B} and Ultraviolet/Optical Telescope (UVOT) \citep{roming2005}. It observes the sources in the sky at hard X-ray bands, soft X-ray bands, ultraviolet, and optical bands. \emph{Swift}-UVOT provides Optical/UV data with V, B, U, W1, M2, and W2 filters. Notably, during most \emph{Astro}Sat observations of Mrk\,501, \emph{Swift} also observed the source simultaneously. The \emph{Swift}-UVOT data of Mrk\,501 are available in all six filters, except during the period MJD 57614 -- 57615, where only V-filter observations are available. 
We retrieved the data of Mrk\,501 from HEASARC Archive and processed it into scientific products using the HEASoft packages namely \textit{UVOTSOURCE} and \textit{UVOTIMSUM}. The \textit{UVOTSOURCE}  package was used to process the images, while the \textit{UVOTIMSUM} package was used to add multiple images in the filters.
The photometry of the source is done by following the instructions from \citet{poole2008}. A  source region is chosen as a circular region of 5 arcsecs centered at the source location, while the background region is selected away from the source location with an area 3 times larger than the source region. The observed flux was corrected for galactic extinction by using $E(V-B) = 0.017$ and  $Rv=Av/E-B = 3.1$,  following  \citep{schlegel1998, schlafly2011}.\\
 
\subsection{\emph{Fermi}-LAT}
The \emph{Fermi} Large Area Telescope (LAT) is a space-based high energy telescope with a wide field of view $\sim$ 2.4 Sr \citep{2009ApJ...697.1071A}. It is part of the \emph{Fermi} $\gamma$-ray space Telescope, launched by NASA in 2008. \emph{Fermi}-LAT operates primarily in scanning mode, monitoring the entire sky in the energy range of approximately 20\,MeV to 500\,GeV every three hours. We acquired the \emph{Fermi}-LAT $\gamma$-ray data of Mrk\,501 during the time period MJD 57615.2 -- 59666. The data was processed using Fermitools version 2.2.0, following standard analysis procedures outlined in \emph{Fermi}-LAT documentation  \citep{2017ICRC...35..824W}. Photon events with a high probability were extracted specifically within a $15'$ region of Interest (ROI) centered at the source location. The analysis was conducted within the energy range of 0.1 -- 300 GeV, employing the “$\rm P8R3_-SOURCE_-V3$” instrument response function (IRF) with specified parameters evclass=128 and evtype=3. XML files were generated utilizing the galactic diffusion model “$\rm gll_-iem_- v07.fits$” and the isotropic background model “$\rm iso_-P8R3_-SOURCE_-V3_-v1.txt$”. Additionally, spectral models and parameters for sources within the ROI were obtained from the fourth \emph{Fermi} source catalog. Subsequently, flux points and corresponding energies from \textit{Fermi}-LAT observations were converted to a format compatible with XSPEC (PHA format) using the “\textit{ftflx2xsp}” tool.

\begin{table*}
 \caption{\emph{Astro}Sat/\emph{Swift-UVOT} observation details of Mrk\,501. Column 1: Observation,   2:  Observation ID, 3: Instrument,  4: Date (MJD),  5: Mode of receiving data,  6: Exposure time.}   

\label{tab:1}
\centering
\begin{tabular}{c c c c c c c}
\hline

Observation &Observation ID &Instrument&Date(MJD) & Mode & Exposure Time(ks)\\
(1) & (2) & (3) & (4) & (5) & (6)\\
\hline
\\
S1&G05 218T05 9000000602 & \emph{Astro}Sat & (57615.2 -- 57615.9)& PC & 29.7\\
&00092398007 & \emph{Swift}-UVOT    & ( 57614.2 ) &PC& 0.96\\
\\
S2&A05 107T01 9000002852 & \emph{Astro}Sat & (58594.9 -- 58596.2) &PC&30\\
&00095331002 &  \emph{Swift}-UVOT & ( 58594.0 )& PC &0.81\\
\\
S3&A07 101T02 9000003544 & \emph{Astro}Sat & (58910.3 -- 58912) &  PC&34.3\\
&00095341008 & \emph{Swift}-UVOT    &  ( 58914.8 ) & PC&0.87\\
\\
S4&A07 145T01 90000003594 & \emph{Astro}Sat & (58934.5 -- 58939.8) &PC &106\\
&00095341012&  \emph{Swift}-UVOT & ( 58942.3 )  &PC & 0.51\\
\\
S5&T05 015T01 9000005026 & \emph{Astro}Sat&(59662.3 -- 59665.8)&PC&82.8\\
&00096029018&\emph{Swift}-UVOT &( 59662.3 )&PC&1.03\\
&00011184185&\emph{Swift}-UVOT &( 59663.10 )&PC& 0.81\\
&00011184186&\emph{Swift}-UVOT&( 59664.1 )&PC& 1.07\\
&00011184187&\emph{Swift}-UVOT&( 59665.1 )&PC& 1.06\\
\hline

\end{tabular}
\end{table*}
\section{\emph{Astro}Sat Results}
\label{astr:res}
\subsection{Temporal Analysis}
\label{temp:2}
To explore the X-ray variability of Mrk\,501, we utilized simultaneous observations of SXT and LAXPC instruments onboard \emph{Astro}Sat during the period  MJD 57615 -- 59666. Mrk\,501 were observed five times during the integration period, with observation details provided in Table \ref{tab:1}.
We obtained 100-second binned SXT and LAXPC  light curve for each observation (see Figure ~\ref{fig:sxt,laxpc}). The resulting light curves were analyzed using the LCSTATS tool, leading to the determination of RMS fractional variation values,  which are mentioned in Table \ref{tab:3}. The obtained RMS fractional values indicate subtle variability in the source. Moreover, the maximum count rate acquired in all the observations  (see Table \ref{tab:3}) suggests that the source remained mostly in a low flux state.
These observations provide an opportunity to investigate Mrk\,501 during its low flux states, facilitating an enhanced understanding of the underlying emission processes.

\begin{figure*}
\centering
\includegraphics[scale=0.3]{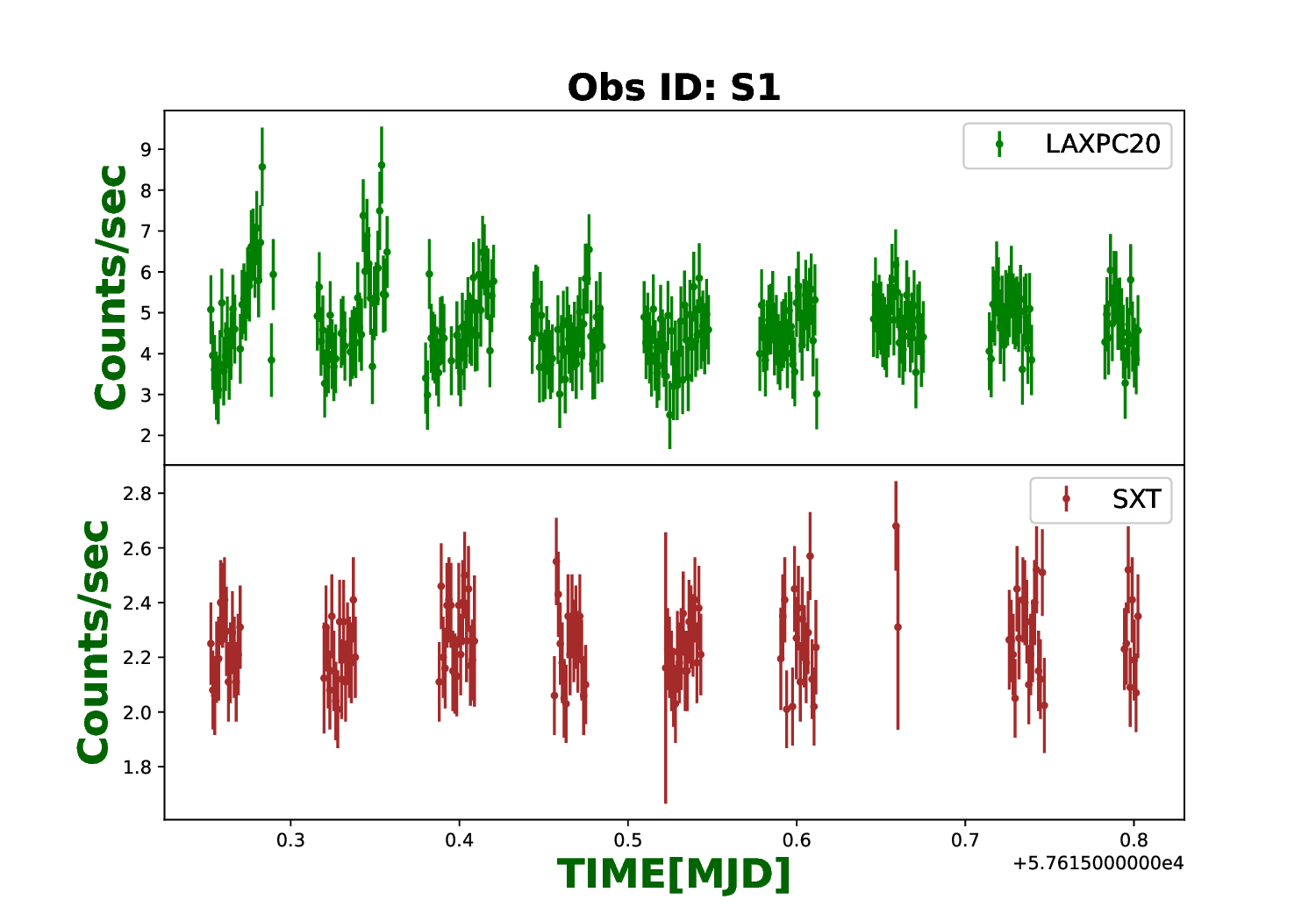}
\includegraphics[scale=0.3]{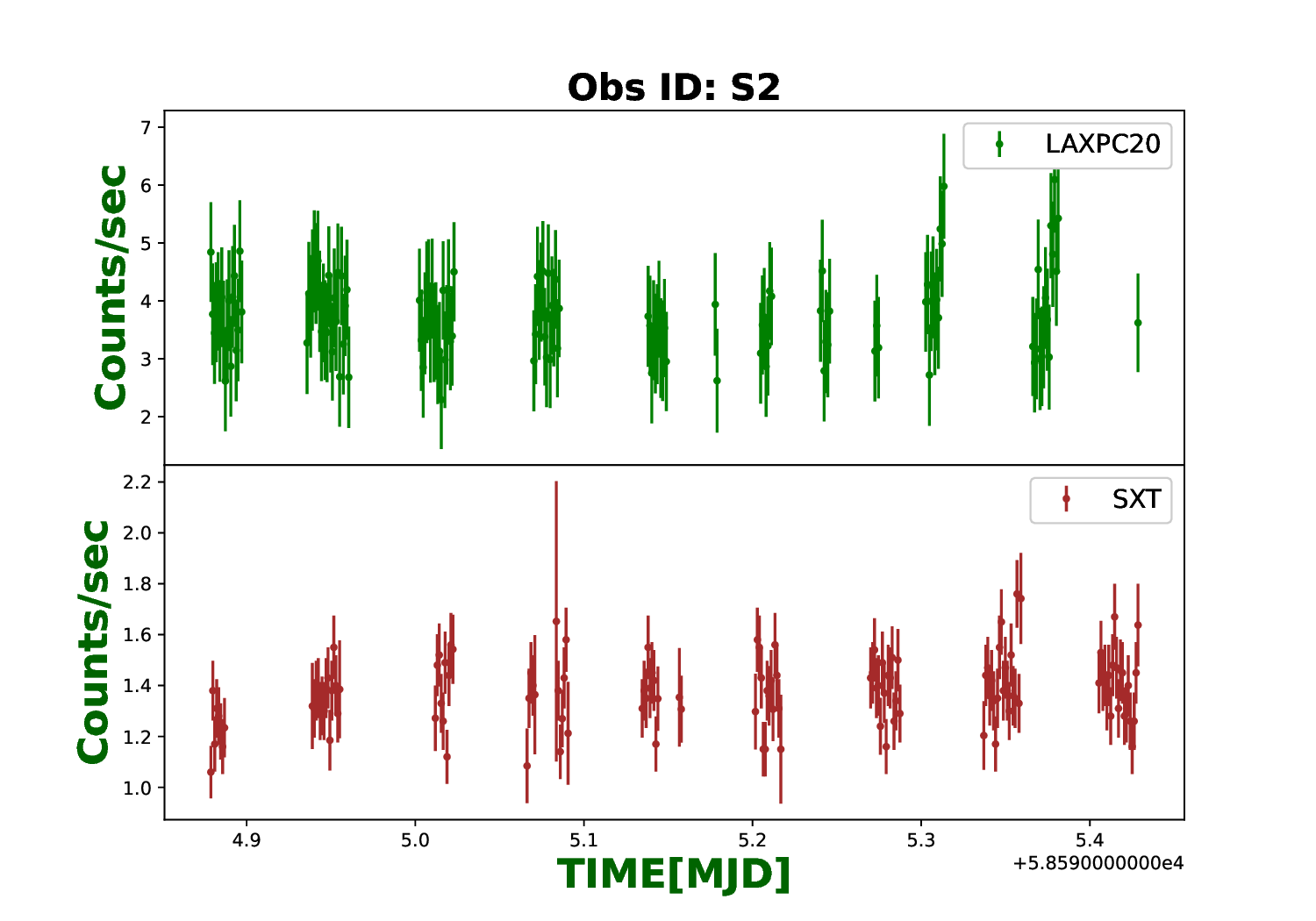}
\includegraphics[scale=0.3]{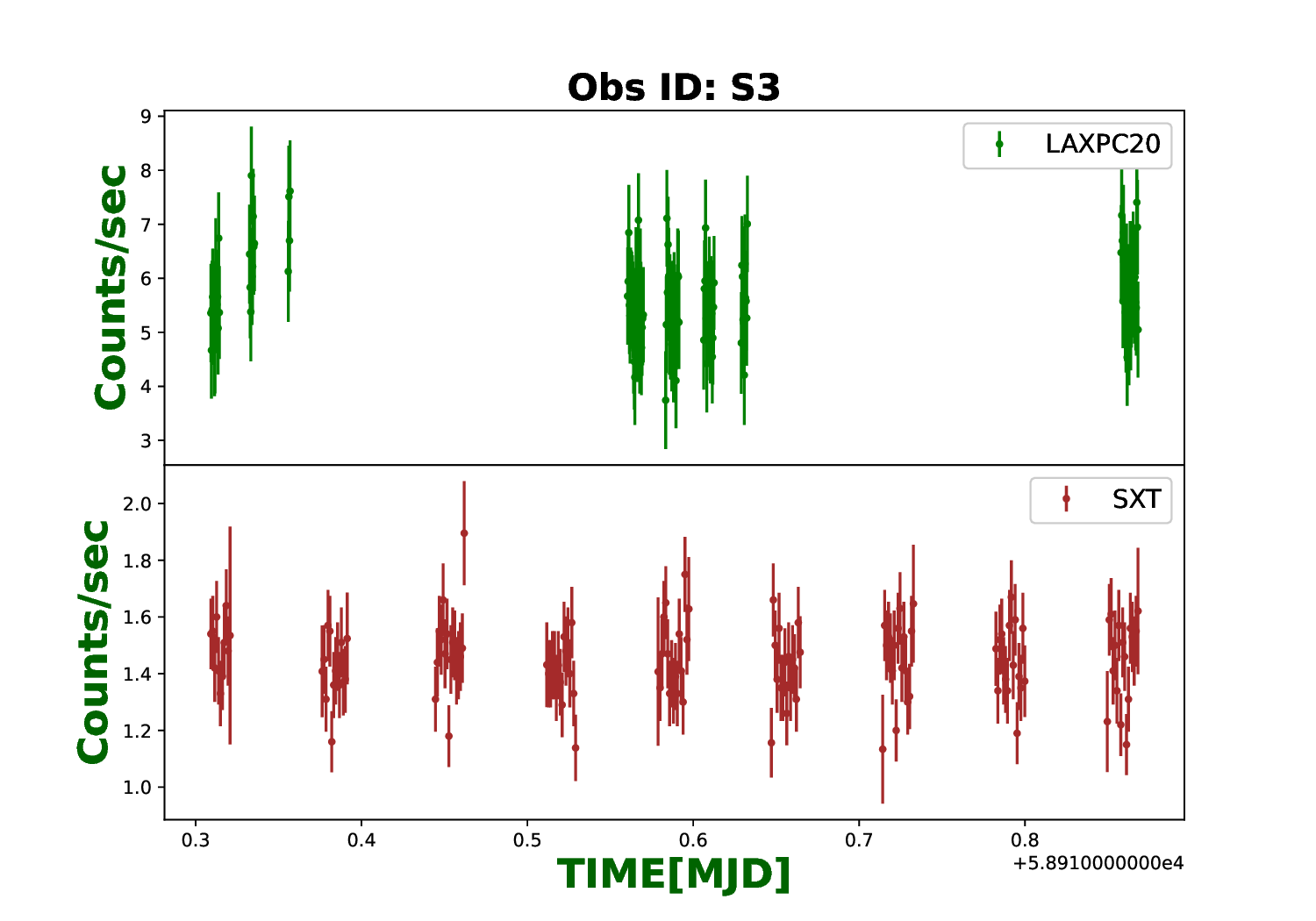}%
\includegraphics[scale=0.3]{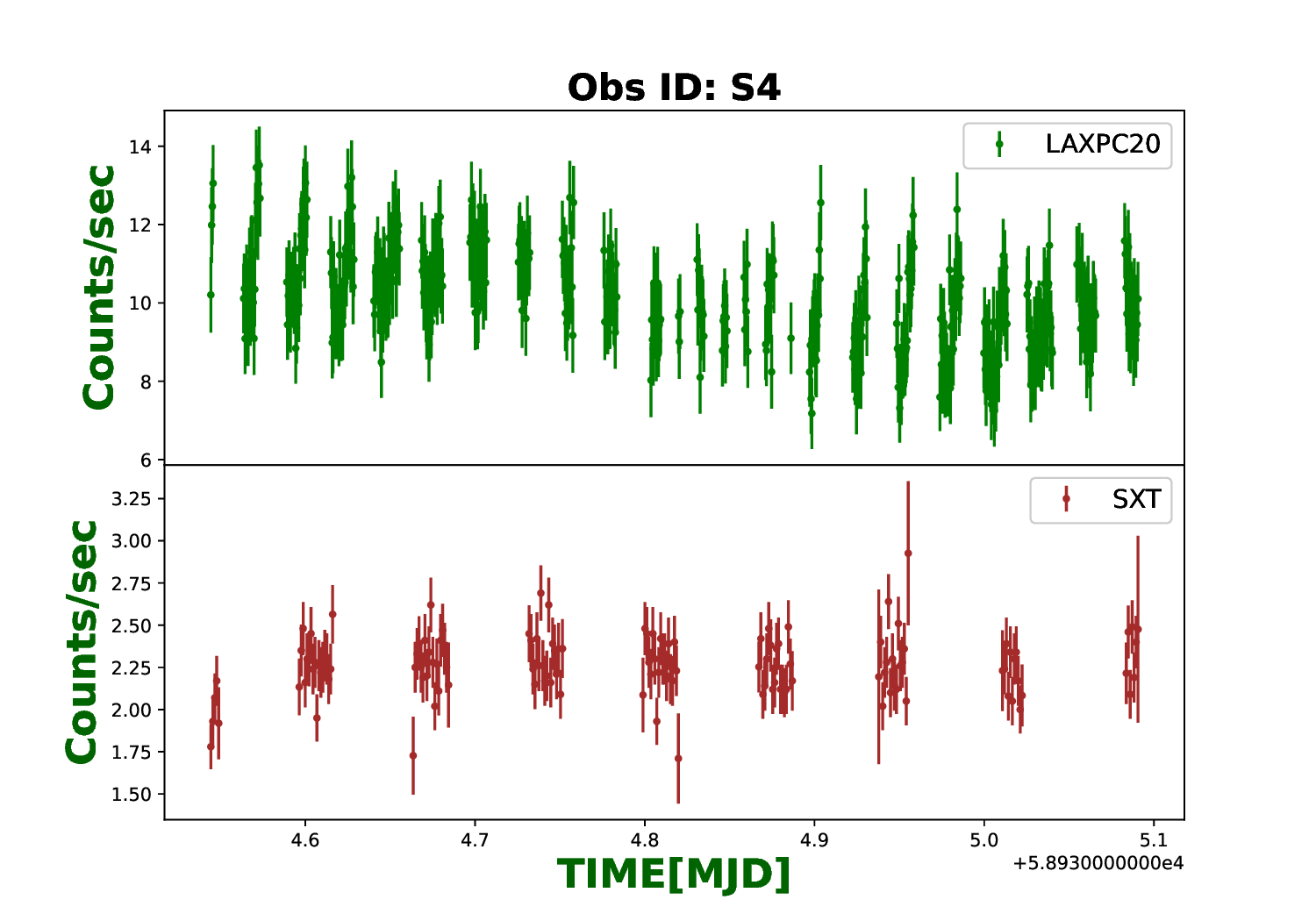}
\includegraphics[scale=0.3]{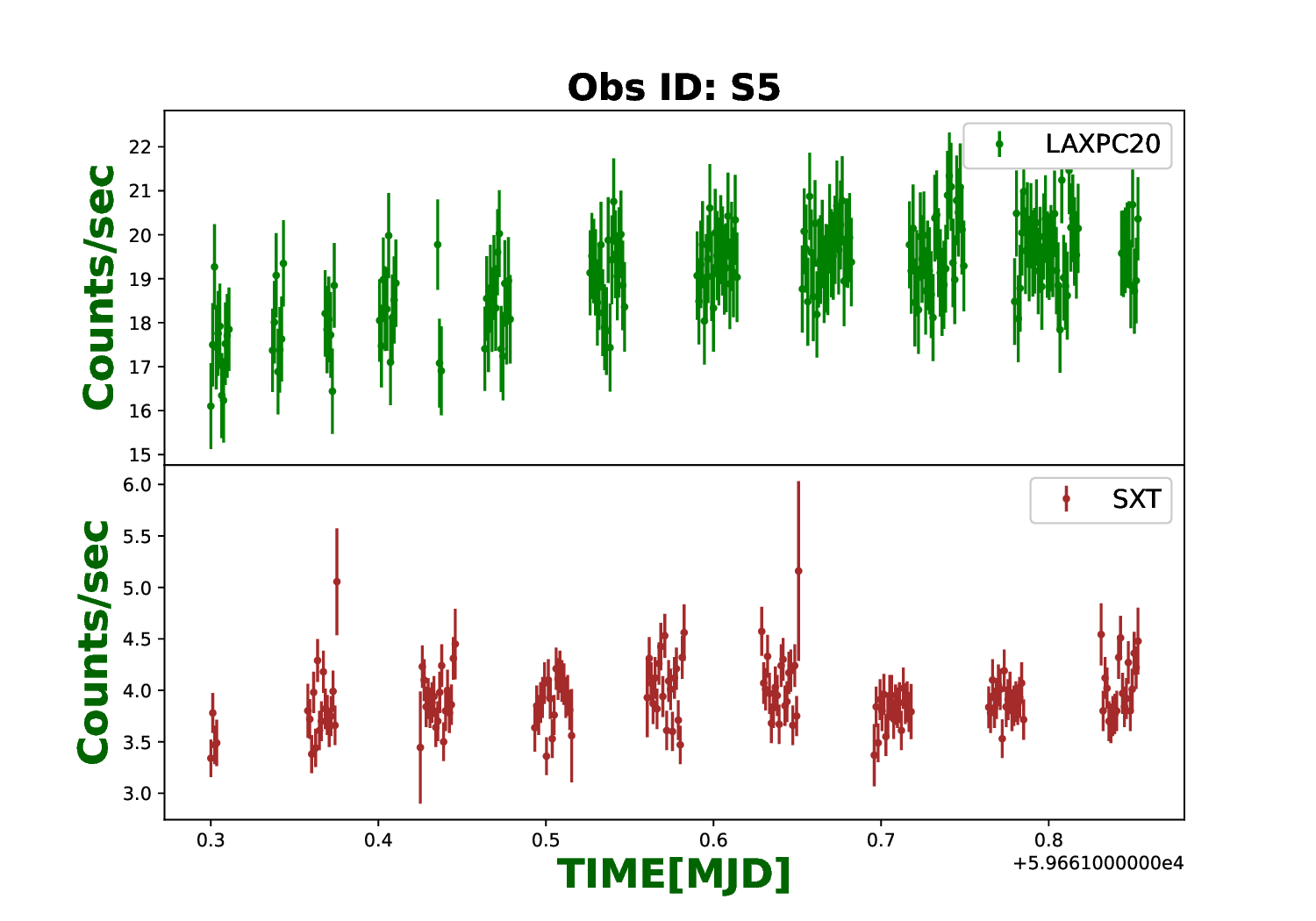}

\caption{Simultaneous 100 second binned SXT  (0.3 -- 7.0\,keV)  and LAXPC (3.0 -- 30.0\,keV) light curves of Mrk\,501. Top left S1 observation, top right  S2 observation, middle left S3 observation, middle right S4 observation, and bottom S5 observation.}
\label{fig:sxt,laxpc}
\end{figure*}

\begin{table*} 
\caption{Details of the  LCSTATS analysis done on the  SXT (0.3 -- 7.0\,keV) and LAXPC (3.0 -- 30.0\,keV) light curves. Columns  1:  Observation, 2: Instrument, 3:  Counts/sec, 4:  RMS fractional variability.}
\centering
\begin{tabular}{c c c c c c }
\hline
Observation  & Instrument & (Counts/sec)& $F_{var}$  \\ \hline   \\
S1 & SXT  & 2.68 $\pm$0.16 & 0.06 $\pm$0.007\\
& LAXPC  & 8.61$\pm$0.90 & 0.07$\pm$0.02\\
\\
S2&SXT&1.76$\pm$0.13&0.07$\pm$0.01\\
& LAXPC & 6.09$\pm$0.91 & 0.08$\pm$0.04\\
\\
S3&SXT&1.89$\pm$0.18&0.03$\pm$0.08\\
&LAXPC&7.90$\pm$0.90&0.04$\pm$0.07\\
\\
S4&SXT & 2.92$\pm$0.42 & 0.08$\pm$0.003\\
&LAXPC&13.51$\pm$0.98&0.09$\pm$0.005 \\
\\
S5&SXT& 5.16$\pm$0.87&0.03$\pm$0.006 \\
&LAXPC&21.46$\pm$0.91&0.04$\pm$0.003\\
\\
\hline
\end{tabular}
\label{tab:3}

  \end{table*}

\subsection{Spectral Analysis}

In order to understand the X-ray spectral behavior of Mrk\,501, we conducted a  joint spectral fitting of  SXT and LAXPC spectra obtained in the energy range  0.3 -- 6.0\,keV and 3.0 -- 30.0\, keV, respectively, for all the five selected observations.
 Further, to account for the absorption due to neutral hydrogen, we used the galactic column density value as  $n_H=1.69\times10^{20} cm^{-2}$ \citep{kalberla2005}.

\begin{figure*}
\centering
\includegraphics[width=.8\textwidth]{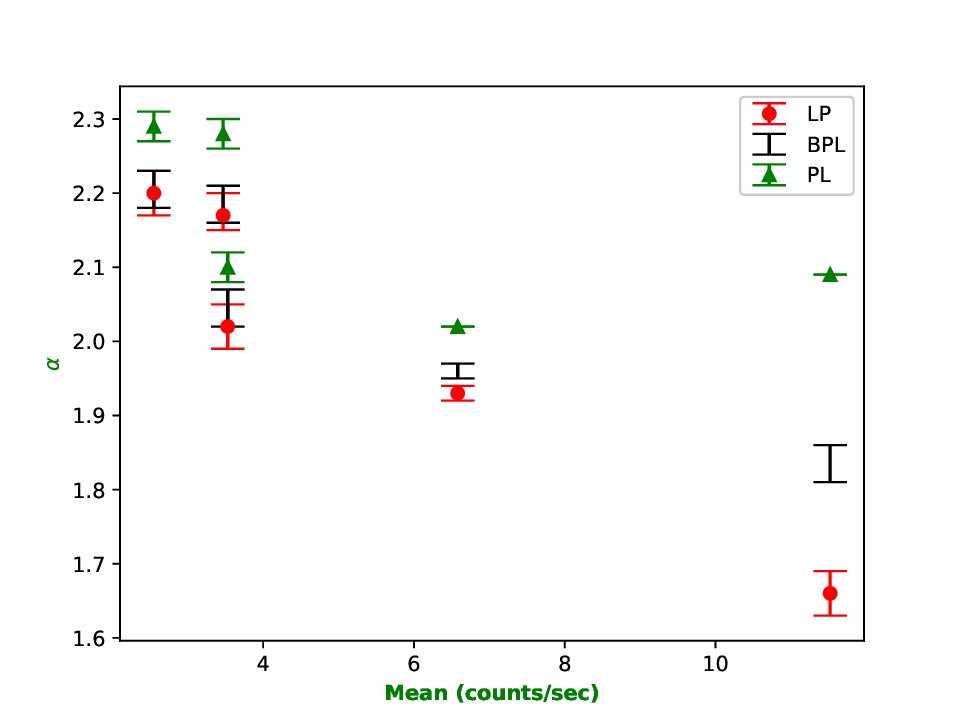}
\caption{ Spectral index versus the mean count rates obtained by fitting the joint SXT-LAXPC spectrum with the LP,  BPL and PL models.}.
\label{fig:meancount}
\end{figure*}

\begin{figure*}
\centering
\includegraphics[scale=0.4,angle=270]{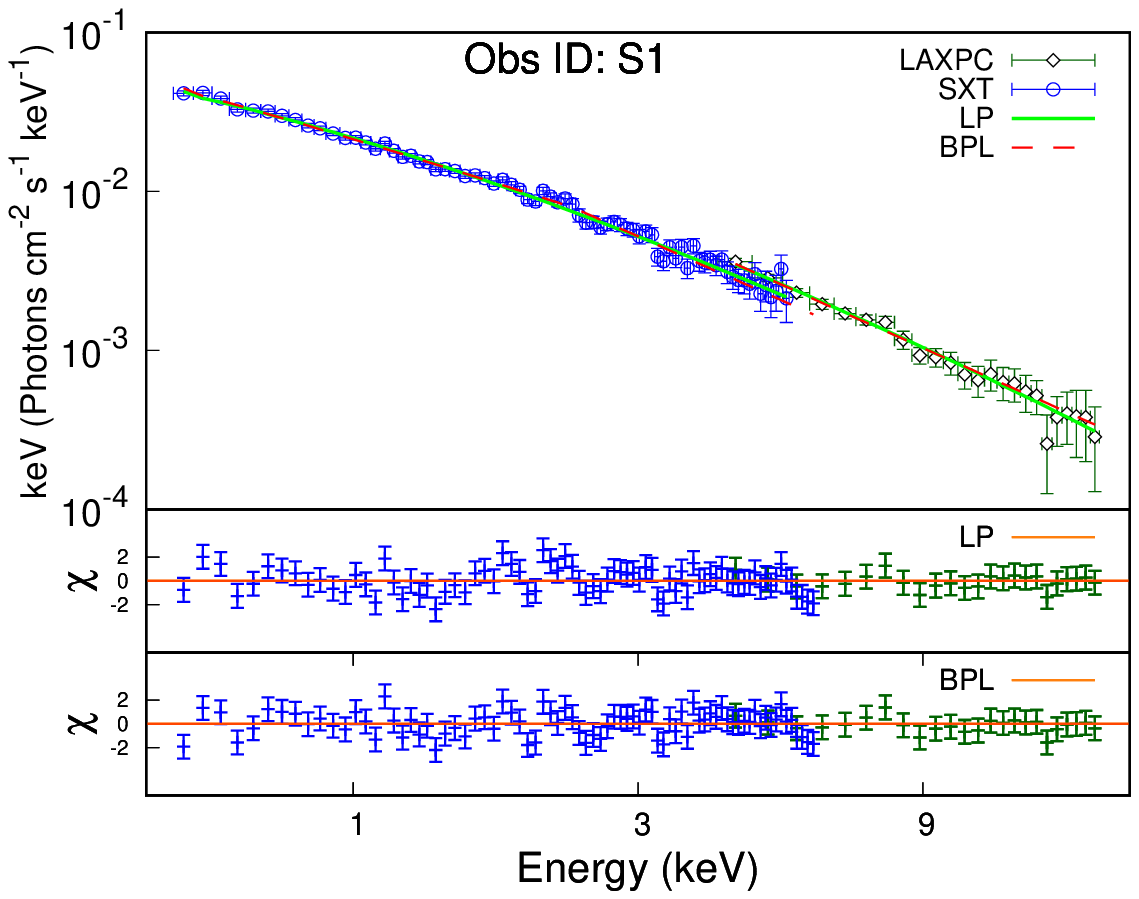}
\includegraphics[scale=0.4,angle=270]{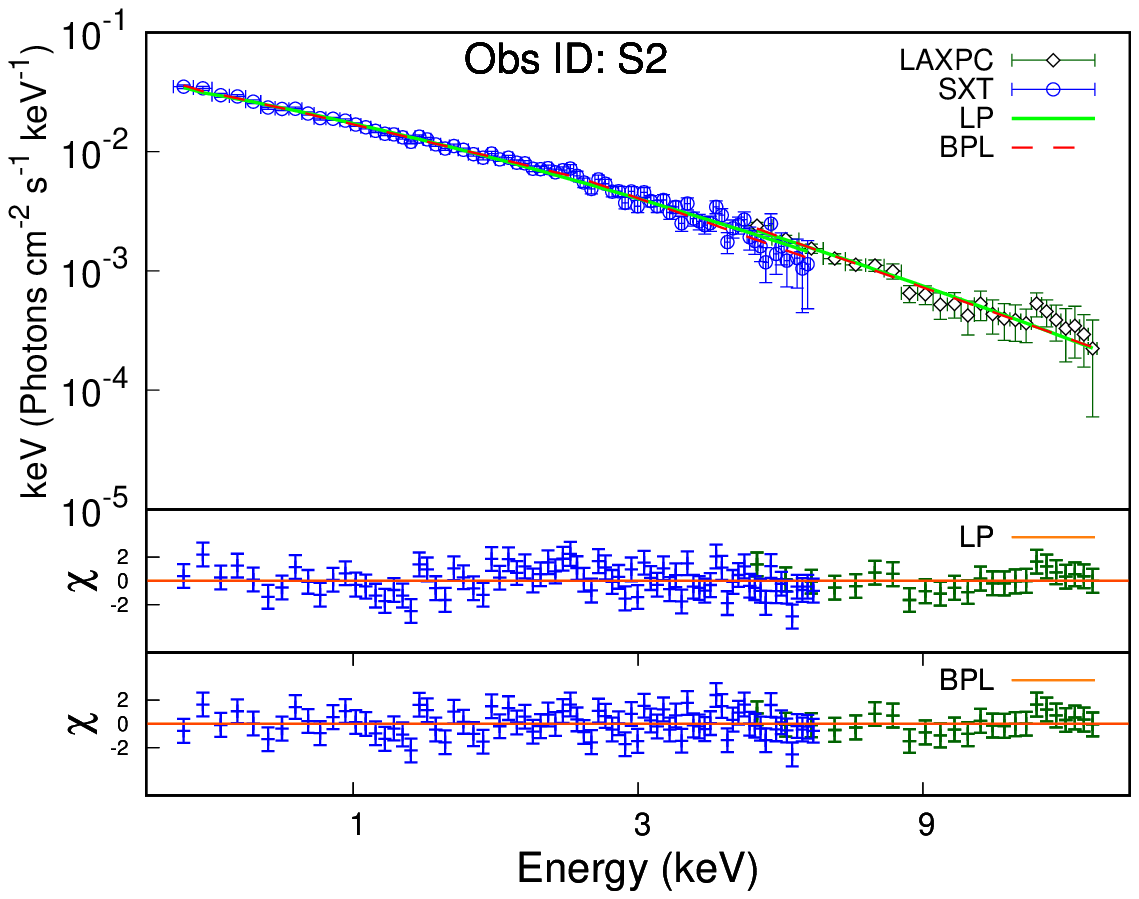}
\includegraphics[scale=0.4,angle=270]{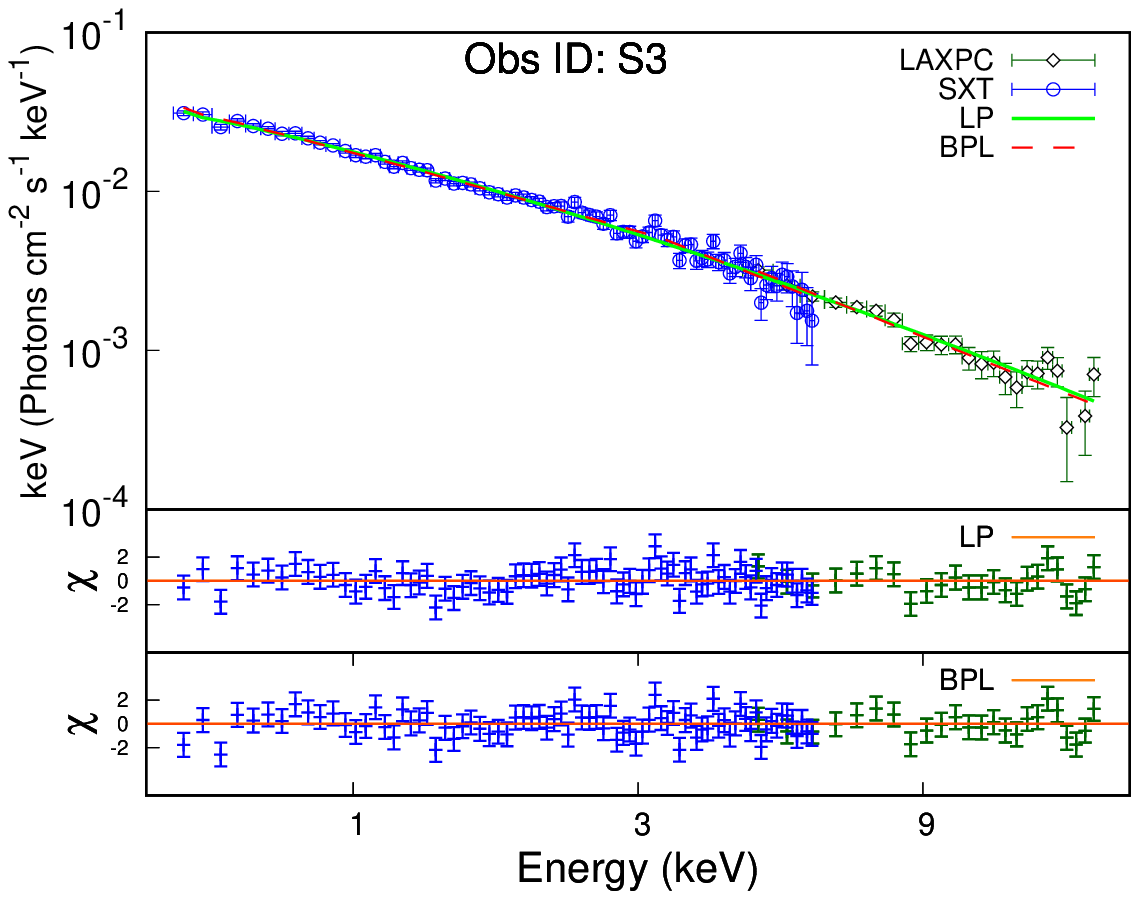}%
\includegraphics[scale=0.4,angle=270]{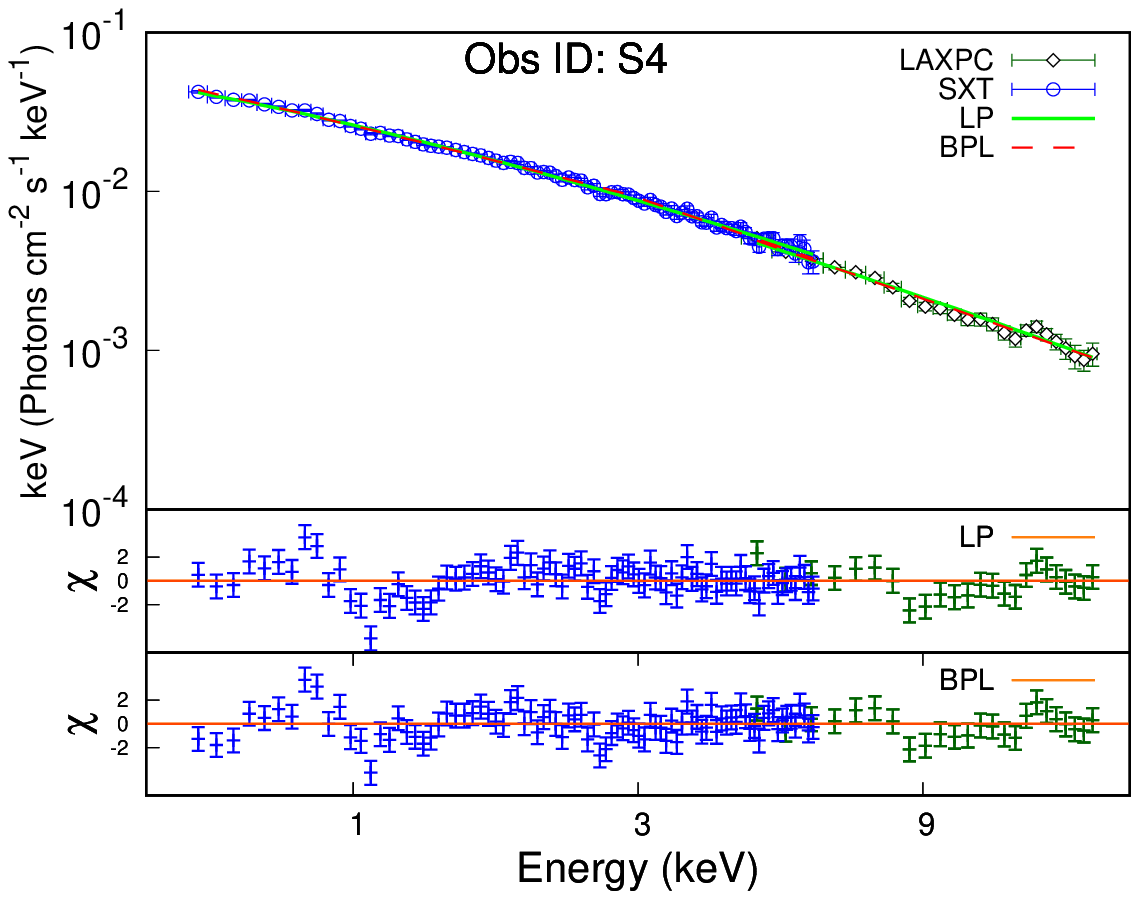}
\includegraphics[scale=0.4,angle=270]{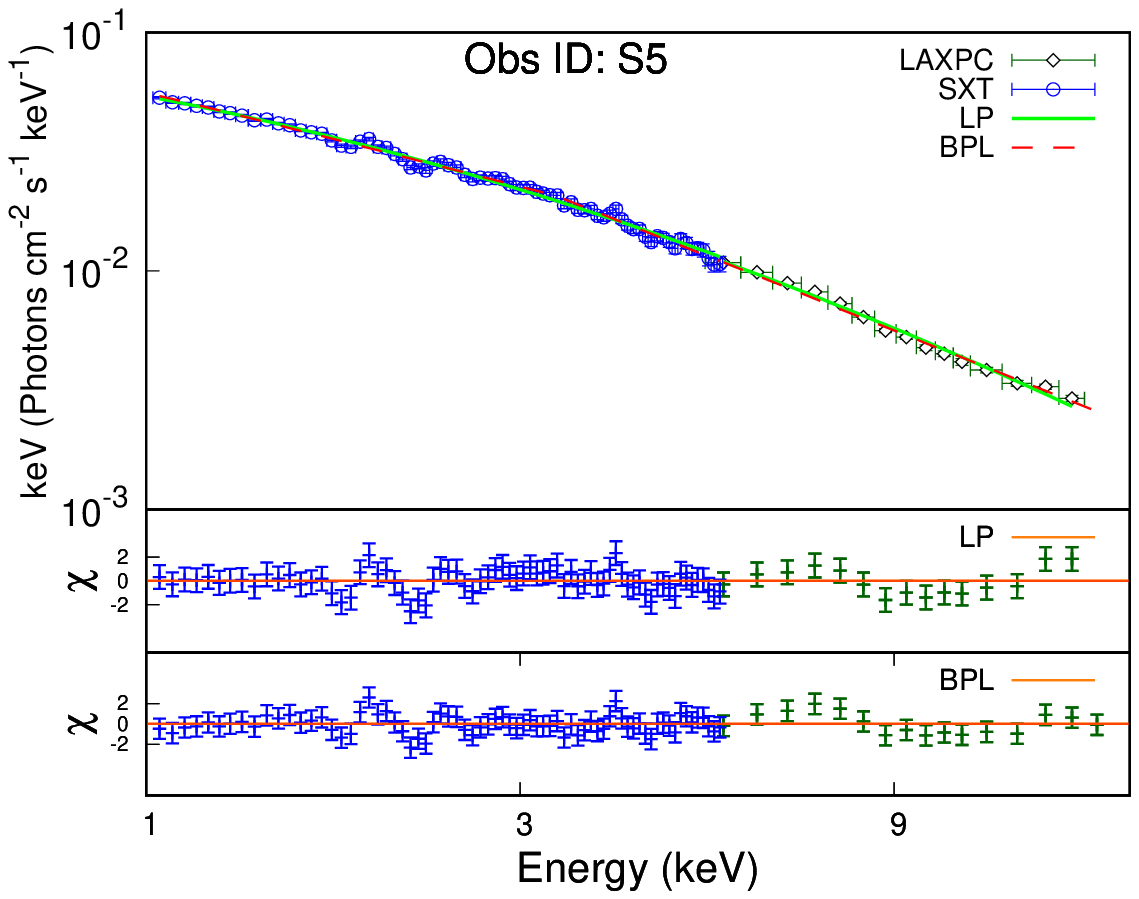}\\
\vspace{12pt}
\caption{The combined SXT and LAXPC spectrum fitted with the constant*TBabs*LP/BPL model. The  five plots correspond to  S1 (top left), S2 (top right), S3 (middle left), S4 (middle right), and S5 (bottom)
observations.}

\label{joint:xray}
\end{figure*}

Three distinct models, namely the power-law (PL), broken power-law (BPL), and logparabola (LP) models, were employed to fit the spectrum. The PL model is defined as
\begin{equation}
    n(E) = K\cdot E^{-\Gamma},
\end{equation}
where K is normalization at 1\,keV, and $\Gamma$ is the photon index. The LP model is defined as
\begin{equation}
\label{eq:2}
n(E)= K (E/E_{pivot})^{-(\alpha+\beta \log(E/E_{pivot}))}
\end{equation}
where $\alpha$ is the photon-index at pivot energy $E_{pivot}= 1$\,keV, $\beta$ is the curvature parameter, and K is the normalization. The BPL model is defined as
\begin{equation}
\label{eq:3}
n(E) =
\begin{cases}
    K E^{-\Gamma_1} & \text{for } E < E_b \\
    K E_b^{\Gamma_2-\Gamma_1} E^{-\Gamma_2} & \text{for } E \geq E_b
\end{cases}
\end{equation}

where $K$ is normalization, $E_{\rm b}$ is break energy, $\Gamma_1$ and $\Gamma_2$ indicate the photon indices before and after break energy $E_\mathrm{b}$. 
The combined broad X-ray spectra were fitted with the $\emph constant\times tbabs\times PL/BPL/LP$ model. The “constant” was considered to take into account any systematic difference in the effective areas of LAXPC and SXT. The best-fit parameters of these models for the five observations are given in Table \ref{tab:a}. As shown in the table, the PL model results in highly reduced-$\chi^2$ values compared to the BPL and LP models. The combined  SXT and LAXPC X-rays spectra fitted with BPL and LP models are shown in Figure~\ref{joint:xray}.  
In the case of BPL, the spectral indices before and after break energy are obtained in the range  1.84 -- 2.21 and 2.30 -- 2.76, respectively with break energy between 2.32 -- 3.33 keV. During the hard spectrum, the source is brighter as compared to other observations.  
In the LP model, the spectral indices denoted as $\alpha$ ranged between 1.66 -- 2.20 and $\beta$ values 0.18 -- 0.36. These values indicate that the spectra of the source exhibit slight to moderate curvature. Notably, the parameters of the LP model are consistent with those of the BPL model, with the $\alpha$ parameter of the LP model being the hardest for the same observation where BPL indices are hardest. Additionally, for the hardest spectrum observation, the curvature parameter shows a maximum value of 0.36. The variation of $\alpha$ with the mean count rate shown in Figure\,\ref{fig:meancount} suggest harder when brighter feature in the source. Similarly, the variations of the spectral indices obtained from the BPL (lower energy index) and PL models with the mean count rates exhibit the same trend, as depicted in the Figure \ref{fig:meancount}.
\\
  

\begin{table*}
\label{tab:2}
 \centering
 \caption{The best-fit parameters obtained by fitting the combined SXT and LAXPC spectrum with the constant*tbabs*BPL/LP/PL models. Columns 1: Observation 2: Model 3: Index before break energy / PL index 4: Index after break energy 5: Index at pivot energy 6: Curvature parameter 7: Norm 8: Constant factor and 9: Reduced-$\chi^2$}

\begin{tabular}{ c c c c c c c c c c}

\hline

Observation &Model&$\Gamma_{1}$& break energy & $\Gamma_{2}$&$\alpha$&$\beta$&norm  &factor& $\chi^2_{red}$ \\ \hline &&&& \\
S1&BPL & $2.19^{+0.02 }_{-0.02}$ & $2.32^{+0.49}_{-0.19}$ &$2.67^{+0.02}_{-0.02}$&&&$0.02^{+0.002}_{-0.002}$ & $0.80^{+0.06}_{-0.06}$ & 90.62/98=0.92 \\ 
\\
&LP&&&&$2.17^{+0.02}_{-0.03}$&$ 0.29 ^{+0.05}_{-0.05}$&$0.02 ^{+0.001}_{-0.001}$&$0.86^{+0.06}_{-0.05}$& 98.35/99=0.99 \\
\\
&PL&$2.28^{+0.02}_{-0.02}$&&&&&$ 0.02^{+0.002}_{-0.001}$&$1.13^{+0.06}_{ -0.05}$&183.64/100=1.83\\
\\
S2&BPL & $2.21^{+0.02 }_{-0.02}$ & $2.42^{+0.45}_{-0.33}$ &$2.76^{+0.10}_{-0.09}$&&&$0.01^{+0.002}_{-0.002}$ & $0.80^{+0.09}_{-0.08}$ & 99.24/99=0.98 \\ 
\\
&LP&&&&$2.20^{+0.03}_{-0.03}$&$ 0.28 ^{+0.06}_{-0.06}$&$0.01 ^{+0.001}_{-0.001}$&$0.91^{+0.08}_{-0.07}$& 119.78/99=0.97 \\
\\
&PL&$2.29^{+0.02}_{-0.02}$&&&&&$ 0.01^{+0.002}_{-0.001 }$&$1.23^{+0.08}_{ -0.07}$&176.69/100=1.76\\
\\
S3&BPL & $2.05^{+0.02}_{-0.02}$ & $3.24^{+0.45}_{-0.33}$ &$2.46^{+0.10}_{-0.09}$&&&$0.01^{+0.001}_{-0.001}$ & $0.93^{+0.07}_{-0.07}$ & 99.24/99=1.00 \\
\\
&LP&&&&$2.02^{+0.03}_{-0.03}$&$0.20^{+0.05}_{-0.05}$&$0.02 ^{+0.001}_{-0.001}$&$1.01^{+0.29}_{-0.46}$& 97.14/100=0.97 \\
\\
&PL&$2.10^{+0.02}_{-0.02}$&&&&&$0.01^{+0.002}_{-0.001 }$&$1.22^{+0.06}_{ -0.06}$&132.14/101=1.30\\
\\
S4&BPL & $1.96^{+0.01}_{-0.01}$ & $2.90^{+0.27}_{-0.21}$ &$2.30^{+0.02}_{-0.02}$&&&$0.02^{+0.001}_{-0.001}$ & $1.02^{+0.04}_{-0.04}$ & 148.26/107=1.38 \\ 
\\
&LP&&&&$1.93^{+0.01}_{-0.01}$&$ 0.18 ^{+0.02}_{-0.02}$&$0.02 ^{+0.001}_{-0.001}$&$1.09^{+0.03}_{-0.04}$& 169.55/108=1.56\\
\\
&PL&$2.02^{-}_{-}$&&&&&$ 0.02^{- }_{-}$&$1.30^{-}_{-}$&316.82/109=2.90\\
\\
S5&BPL & $1.84^{+0.02}_{-0.02}$ & $3.33^{+0.26}_{-0.20}$ &$2.31^{+0.02}_{-0.02}$&&&$0.05^{+0.001}_{-0.001}$ & $1.0^{+0.04}_{-0.04}$ & 67.25/83=0.81 \\ 
\\
&LP&&&&$1.66^{+0.03}_{-0.03}$&$ 0.36^{+0.04}_{-0.04}$&$0.05 ^{+0.002}_{-0.002}$&$1.03^{+0.02}_{-0.02}$& 76.06/83=0.91\\
\\
&PL&$2.09^{-}_{-}$&&&&&$ 0.06^{- }_{-}$&$1.0^{-}_{-}$&486.67/86=5.65\\

\\
\hline
  
\end{tabular}
\label{tab:a}

\end{table*}
\section{Multi-wavelength Analysis} \label{mul:anl}

 \subsection{Multi-wavelength light curve (MWLC)}
\begin{figure*}
\centering
\includegraphics[height=0.3\textheight]{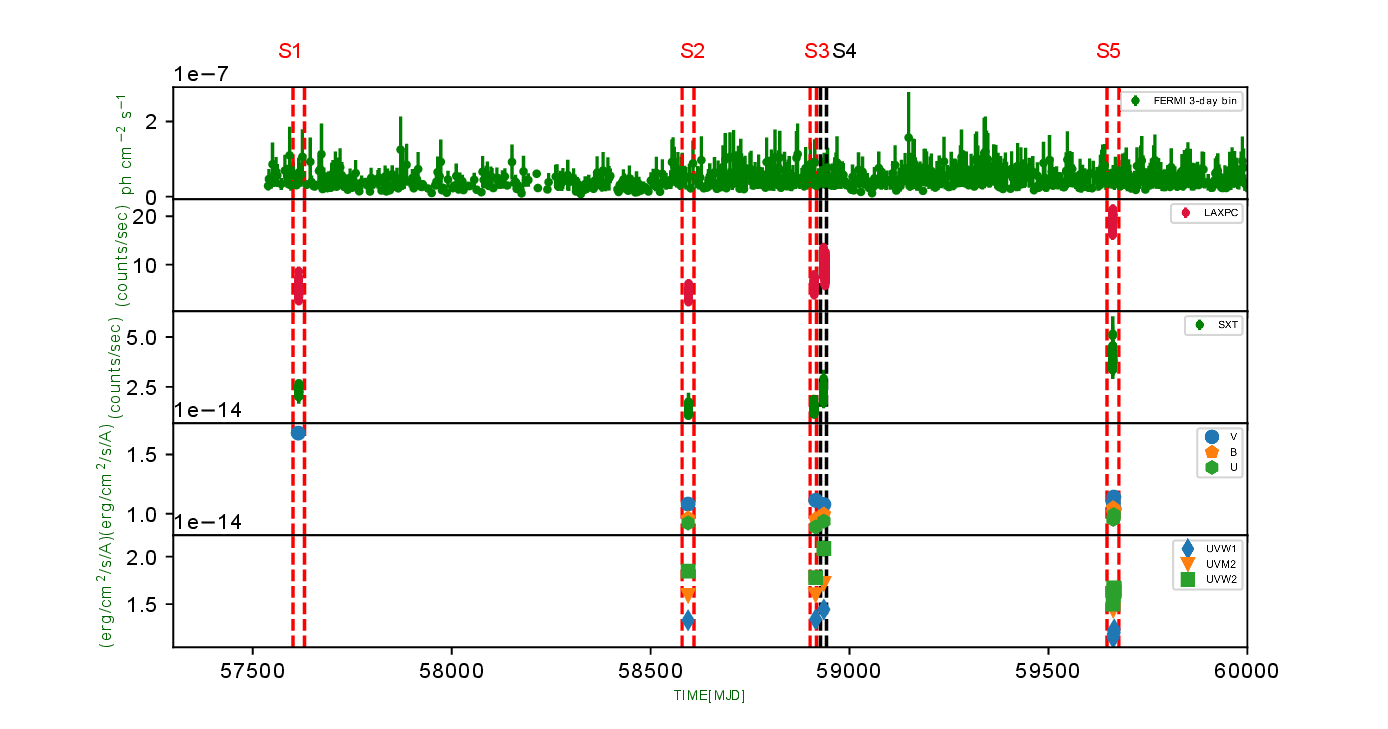}

\caption{Multi-wavelength light curve (MWLC) obtained from \emph{Astro}Sat (SXT, LAXPC), \emph{Swift}-UVOT, and \emph{Fermi}-LAT observations. The 3-day bin $\gamma$-ray light curve, acquired in the energy range of 0.3--300 GeV, is shown in the top panel. The 100-second binned SXT and LAXPC light curves, which were produced by integrating photons in the energy ranges of 0.3 and 7.0 keV and 3.0 and 30.0 keV, respectively, are displayed in the second and third panels.  The optical/UV light curves are shown in final two panels, each point corresponds to individual observation ID. S1, S2, S3, S4 and S5 states represented by dashed vertical lines indicate the time intervals for which broadband spectral modeling is performed.} 

\label{multi_wavelength}
\end{figure*}

 In addition to the \emph{Astro}Sat LAXPC and SXT data, we utilized \emph{Swift}-UVOT and \emph{Fermi}-LAT data observed during the  MJD 57615 -- 59666 to explore the temporal and spectral behavior of source across a broad range of energies. The MWLC obtained using the observations from \emph{Swift}-UVOT, \emph{Astro}Sat-SXT/LAXPC20, and \emph{Fermi}-LAT is shown in Figure~\ref{multi_wavelength}. \emph{Astro}Sat conducted five observations of Mrk\,501 during MJD 57600 -- 59665, but due to limited simultaneous UVIT  observations in the \emph {Astro}browse, we relied on \emph{Swift}-UVOT observations. The MWLC shows that the source has remained mostly in the low flux states. Notably, the MAGIC Collaboration has also reported a historically low flux state in the source during the period 2017 -- 2019 \citep{abe2023multi}. Therefore, \emph{Astro}Sat observations together with \emph{Fermi}-LAT and \emph{Swift}-UVOT data provide an opportunity to carry out a detailed broadband study of the Mrk\,501 in the low flux state.

\subsection{Fractional Variability}
To assess the variability of the source in the low flux state, we computed the fractional variability of the light curves across different energy bands. We employed the procedure outlined in the \citep{2003MNRAS.345.1271V}, wherein the fractional variability $\emph F_{var}$ is determined using the formula 

 \begin{equation}
    \label{eq3}
    F_{var}=\sqrt{\frac{S^2-\overline{\sigma_{err}^2}}{\overline{F}^2}}
\end{equation}
where $\rm S^2$ is the variance, $\rm \overline{F}$ is the mean, and $\rm \overline{\sigma_{err}^2}$ is the mean square of the measurement error on the flux points. The uncertainty on $\rm F_{var}$ is  given by \citet{2003MNRAS.345.1271V}
\begin{equation}
    \label{eq4}
    F_{var,err}=\sqrt{\frac{1}{2N}\left(\frac{\overline{\sigma_{err}^2}}{F_{var}\overline{F}^2}\right)^2+\frac{1}{N}\frac{\overline{\sigma_{err}^2}}{\overline{F}^2}}
\end{equation}
Here, N represents the number of flux points in the light curve. \\

\begin{figure*}
\centering
\includegraphics[]{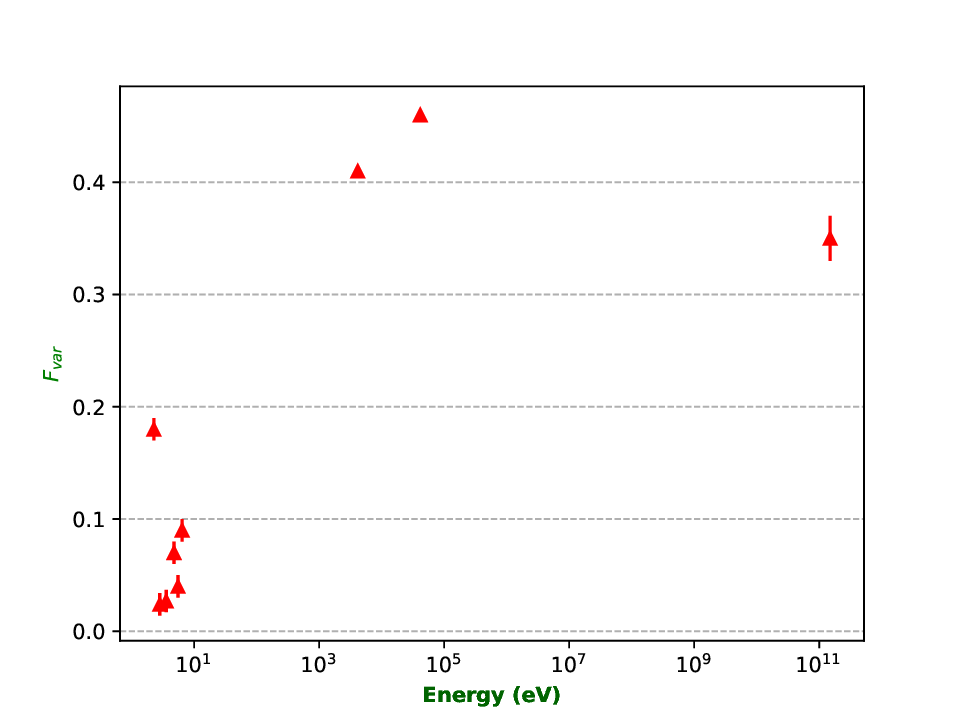}
\caption{Fractional variability amplitude obtained in different energy bands is plotted against the energy.}

\label{fig:fvar}
    
\end{figure*}

 The values of $\emph F_{var}$ along with the uncertainties obtained for $\gamma$-ray, X-ray, and Optical/UV light curves are shown in Table \ref{tab:4}, and the plot between $\rm F_{var}$ and energy is shown in Figure~\ref{fig:fvar}. Notably, the X-ray band exhibits a large variability amplitude value compared to the values obtained in Optical/UV and $\gamma$-ray bands. This variability pattern resembles the characteristic double hump feature observed in the broadband SED of Mrk\,501. The X-ray spectra, lying around the SED peak, are primarily attributed to high-energy electrons, whereas the $\gamma$-ray and optical spectra, positioned before the break energy, arise from low-energy electrons. The faster cooling of high-energy electrons emitting X-rays may lead to a larger variability amplitude at the X-ray band, while the slower cooling of low-energy electrons may result in a smaller variability amplitude at the Optical/UV and $\gamma$-ray band.

\begin{table*}
\centering
\caption{Fractional variability amplitude $F_{var}$ obtained in different energy bands. Columns 1: energy band 2: fractional variability amplitude.}

\begin{tabular}{l r}

\hline
Energy band  & $\rm F_{var}$ \\
\hline
$\gamma$-ray (\mbox{0.1 -- 300\,GeV}) & 0.35$\pm$0.08  \\
LAXPC (3.0 -- 30.0\,keV) &0.46$\pm$0.003\\
SXT (0.3 -- 7.0\,keV) &0.41$\pm$0.002\\
UVW2 & 0.09$\pm$0.05 \\
UVM2 & 0.04$\pm$0.01\\
UVW1 & 0.07$\pm$0.02\\
U & 0.02 $\pm$ 0.01\\
B & 0.02 $\pm$ 0.01\\
V & 0.18 $\pm$ 0.01\\
\hline
\end{tabular}
\label{tab:4}
\end{table*}

 \section{Broadband spectral analysis}
 \label{broad:band}
  These simultaneous observation offers an opportunity to model the emission of the source with enhanced reliability in the low flux states. Notably, previous modeling efforts were mainly based on the high activity state of Mrk\,501. Although the source was previously examined during its low activity state, the simultaneous observations used for modeling were limited \citep{Anderhub_2009}. 
  
 We modeled the broadband SED of Mrk\,501 by considering a one-zone leptonic model involving synchrotron and SSC processes. In the model, we assume that emission originates from a spherical region of radius R filled with a relativistic particle distribution n($\gamma$). The emission region moves down the jet with bulk Lorentz factor $\Gamma$ at a small angle $\theta$ with respect line of sight of an observer. The close pointing of the relativistic jet results in the Doppler-boosted emission, which is taken into account by incorporating the Doppler factor, $\delta=\frac{1}{\Gamma(1-\beta\,cos\theta)}$.
 The relativistic particles in the emission region undergo synchrotron and IC losses in the presence of magnetic fields and seed photons, respectively. In our model, we assume the SSC process, so seed photons for IC process are synchrotron photons from the jet itself. The electron Lorentz factor, $\gamma$ is expressed in terms of new variable $\xi$, such that $\xi=\gamma\sqrt{\mathbb{C}}$, where $\mathbb{C}=1.36 \times 10^{-11}\frac{\delta B}{1+z}$ with z being the redshift of source.
The synchrotron flux recieved by the observer with $\xi$ as parameter can be written as \citep{10.1093/mnras/stad3534} \\
\begin{equation}\label{eq:syn_flux}
 F_{syn}(\epsilon)=\frac{\delta^3(1+z)}{d_L^2} V  \mathbb{A}  \int_{\xi_{min}}^{\xi_{max}} f(\epsilon/\xi^2)n(\xi)d\xi,
 \end{equation}
 where $\rm d_L$  is luminosity distance, V is the volume of emission region, $\rm \mathbb A = \frac{\sqrt{3}\pi e^3 B}{16m_e c^2 \sqrt{\mathbb{C}}}$,  $\xi_{min}$ and $\xi_{max}$ correspond to the minimum and maximum energy of electron,  and f(x) is the synchrotron emmisivity function \citep{1986rpa..book.....R}. The
 SSC flux received by the observer at energy $\epsilon$  can be obtained using the equation 
 \begin{equation}\label{eq:ssc_flux}
  \begin{split}
 F_{ssc}(\epsilon) =\frac{\delta^3(1+z)}{d_L^2} V  \mathbb{B} \epsilon & \int_{\xi_{min}}^{\xi_{max}} \frac{1}{\xi^2}  \int_{x_1}^{x_2}   \frac{I_{syn}(\epsilon_i)}{\epsilon_i^2}  \\
&  f(\epsilon_i, \epsilon, \xi/\sqrt{\mathbb{C}}) d\epsilon_i   n(\xi)d\xi
 \end{split}
 \end{equation}
where, $\rm \epsilon_i$ is incident photon energy,   $\rm \mathbb{B} = \frac{3}{4}\sigma_T\sqrt{\mathbb{C}}$,  $\rm I_{syn}(\epsilon_i)$ is the synchrotron intensity,  $\rm x_1=\frac{\mathbb{C} \, \epsilon}{4\xi^2(1-\sqrt{\mathbb{C}} \,\epsilon/\xi m_ec^2)}$,  $\rm x_2=\frac{\epsilon}{(1-\sqrt{\mathbb{C}}\,\epsilon/\xi m_e c^2)}$ and

\begin{equation}
f(\epsilon_i, \epsilon, \xi)= 2q\log q+ (1+2q)(1-q)+\frac{\kappa^2q^2(1-q)}{2(1+\kappa q)} \nonumber
\end{equation}
here $\rm q=\frac{\mathbb{C}\epsilon}{4\xi^2\epsilon_i(1-\sqrt{\mathbb{C}}\epsilon/\xi m_ec^2)}$ and $\rm \kappa=\frac{4\xi\epsilon_i}{\sqrt{\mathbb{C}} m_e c^2}$.\\

The Equations \ref{eq:syn_flux} and   \ref{eq:ssc_flux} are solved numerically and the resultant numerical code is incorporated as a local convolution model in XSPEC. In the convolved  XSPEC model `energy' variable is interpreted as $\xi = \gamma \sqrt{\mathbb{C}}$. The convolution model enables the statistical fitting of broadband SED and allows us to model the broadband spectrum for any particle energy distribution $n(\xi)$.   
In our work, we considered different forms of $n(\xi)$ like BPL, LP, energy-dependent acceleration (EDA) model, and particle distribution with maximum energy ($\xi_{max}$) model to fit the broadband SED. Aforementioned, particle distributions like LP and BPL are defined in equations \ref{eq:2} and \ref{eq:3}, respectively. For the $\rm \xi_{max}$ model, we considered the scenario in which shock-induced particle acceleration leads to radiative energy loss. 
Given that the radiative loss is proportional to the square of energy, the loss becomes more significant at higher energies, consequently shaping the particle distribution with maximum energy. The form of particle energy distribution can be written as (after transforming $\gamma$ to $\xi = \sqrt{\mathbb{C}} \gamma$)  \citep{Hota:2021csa}

\begin{equation}
n(\xi) = K \xi^{-p} \left(1-\frac{\xi}{\xi_{max}}\right)^{(p-2)}
\end{equation}

where $K$ is particle normalization, $p$  is the particle spectral index and $\xi_{max}=\gamma_{max}\sqrt{\mathbb{C}}$ is Lorentz factor corresponding to maximum particle energy.  

We have also taken into consideration a scenario in which the acceleration process is energy dependent through the relation $\tau_{acc}=\tau_{acc,R}\left(\frac{\gamma}{\gamma_R}\right)^\kappa$, where $\tau_{acc}$ is the acceleration time scale, $\gamma_R$ is the Lorentz factor corresponding to a gyration radius that matches with the size emission region. The situation can be visualized by assuming that the electrons gain relativistic energy by crossing a shock front and this acceleration ceases when they diffuse away from the shock. The particle distribution for this case can be obtained as \citep{Hota:2021csa}

 \begin{equation}
	n(\xi)=K \xi^{\kappa-1}\rm{exp}\left[-\frac{\psi}{\kappa}\,\xi^{\kappa}\right]
\end{equation}
where  K is normalizartion, $\psi$ is parameter which depends on $\tau_{esc}$ and $\kappa$ through relation $\psi=\tau_{acc,R}/\tau_{esc}(\mathbb{C}\gamma_R^2)^{-\kappa/2}$.

 Using the convolved SED model involving synchrotron and SSC processes, the observed broadband spectrum is determined mainly by parameters such as N,  $\Gamma$, B, R, $\theta$, $\rm \xi_{min}$, $\rm \xi_{max}$ and alongside parameters related to the underlying particle distribution. For example, if the underlying particle distribution is EDA model, the additional parameters include $\kappa$  and $\psi$. The convolution broadband SED code also enables the fitting of the SED with jet power $P_{jet}$ as one of the parameters; however, in this scenario, N must remain a fixed parameter.
 
 By adopting the conventional method, which assumes the inertia of the jet is mainly contributed by cold protons, equal in number to non-thermal electrons, we estimated the total jet power as  \citep{10.1111/j.1365-2966.2007.12758.x, Ghisellini_2014}
\begin{equation}
P_{\text{jet}} = \pi R^2 \Gamma^2 \beta c (U_e + U_p + U_B)
\end{equation}
where 
$\Gamma$  is the bulk Lorentz factor of the jet,  $\beta  = \frac{v}{c}$  is the velocity of the jet in units of the speed of light, R  is the radius of the jet, 
 $U_e$, $U_p$, and $U_B$ are the energy density of the electrons, protons, and the magnetic field repectively.

\begin{figure*}
\centering
\includegraphics[scale=0.3,angle=270]{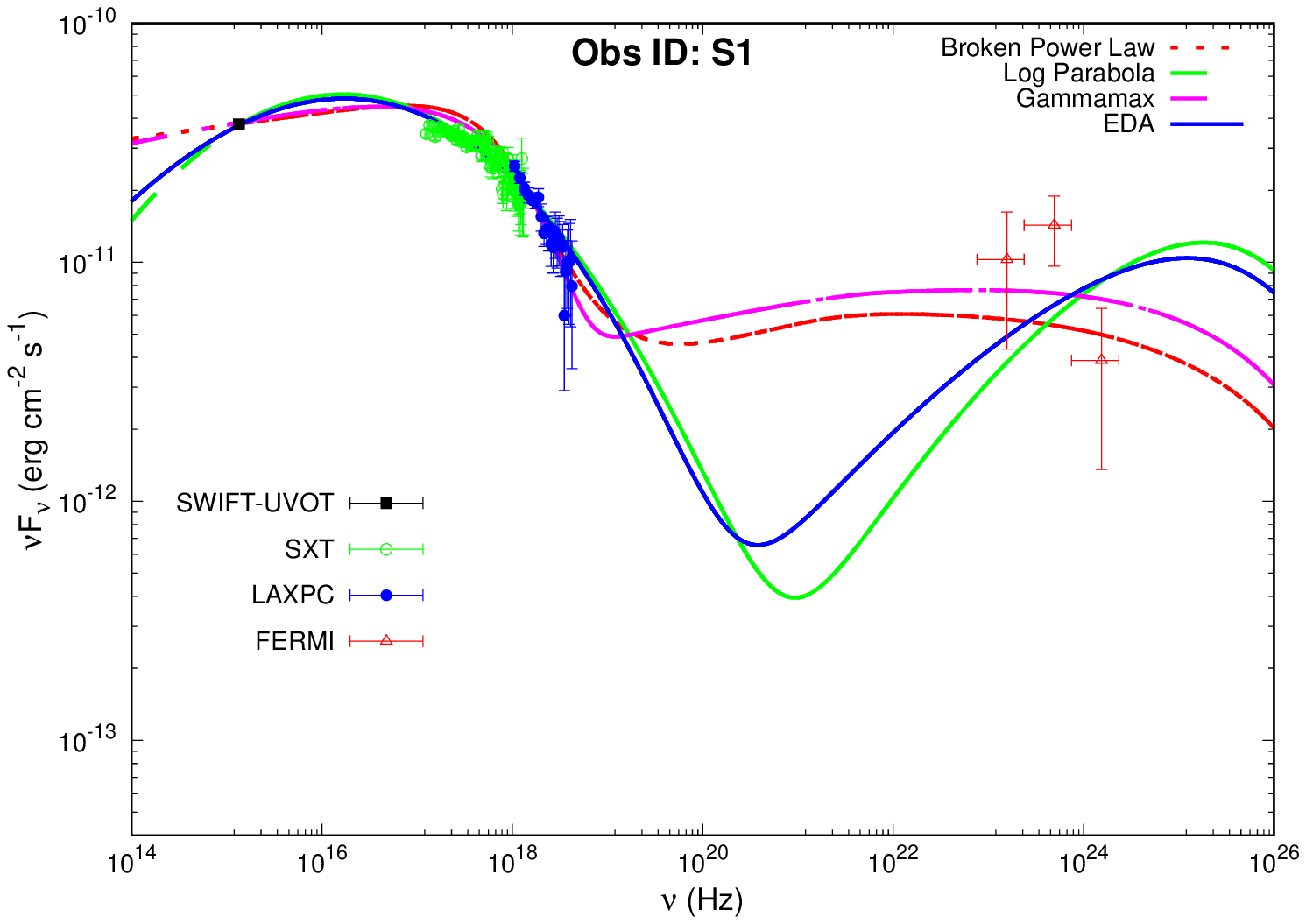}%
\includegraphics[scale=0.3,angle=270]{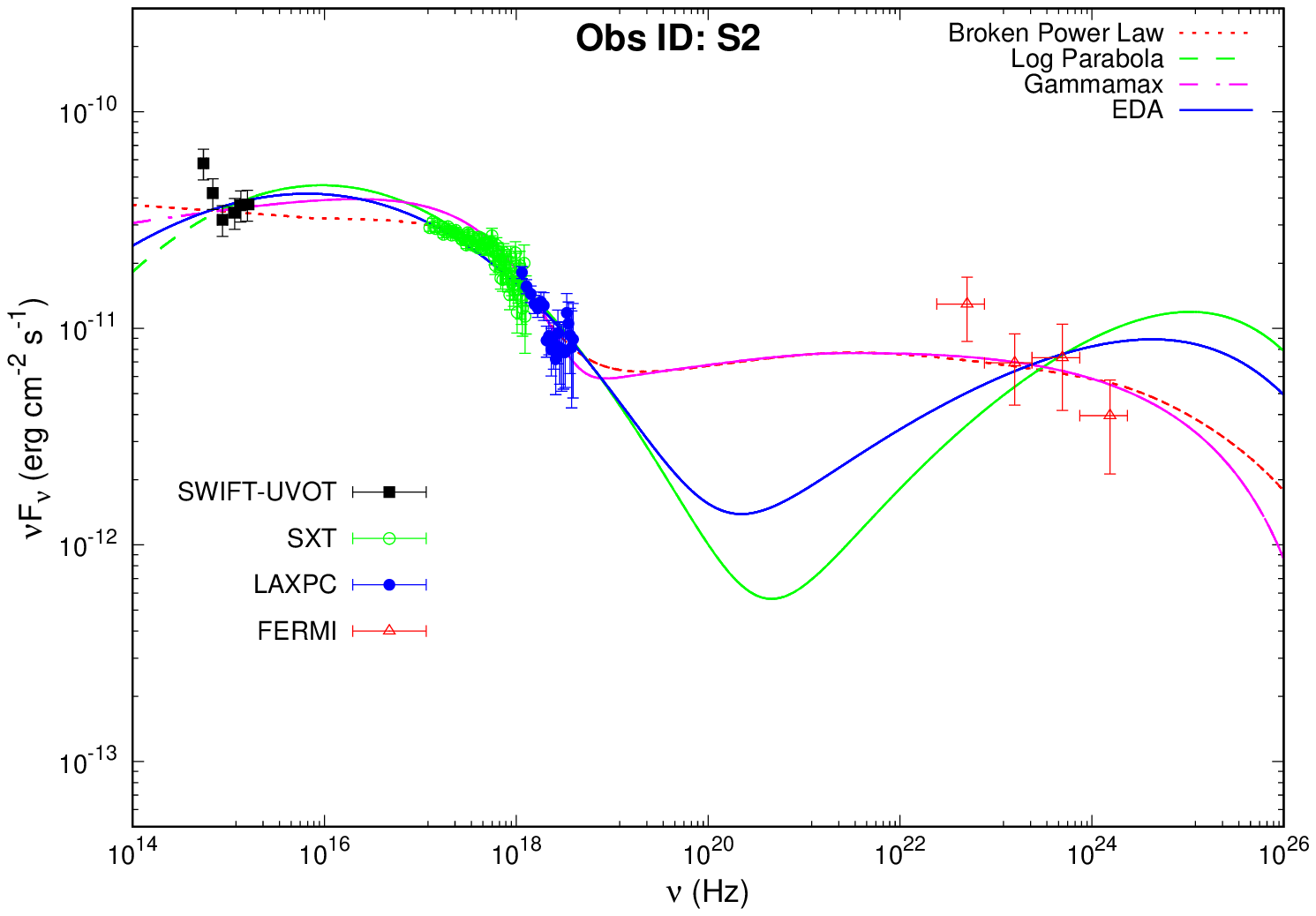}\\

\includegraphics[scale=0.3,angle=270]{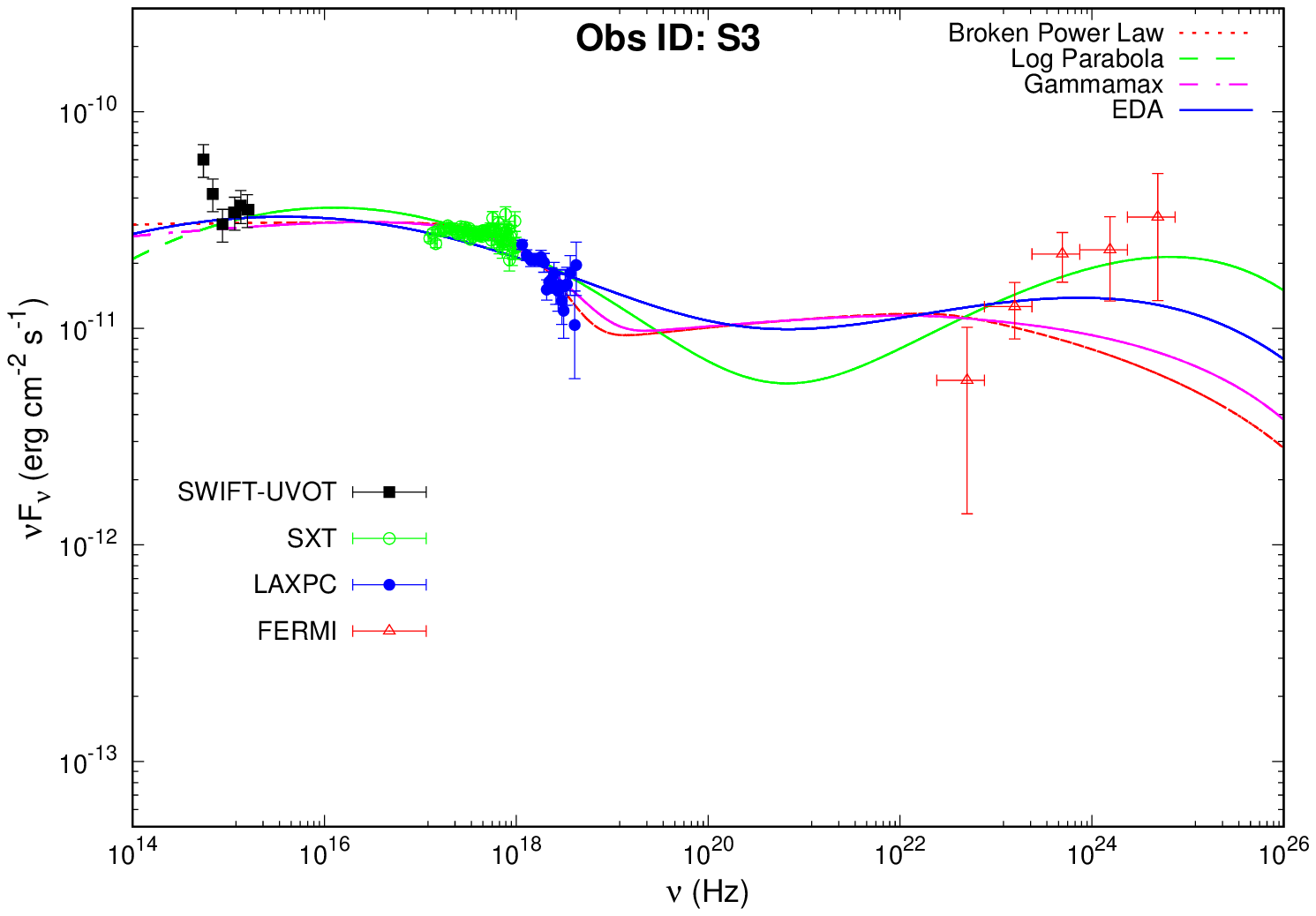}
\includegraphics[scale=0.3,angle=270]{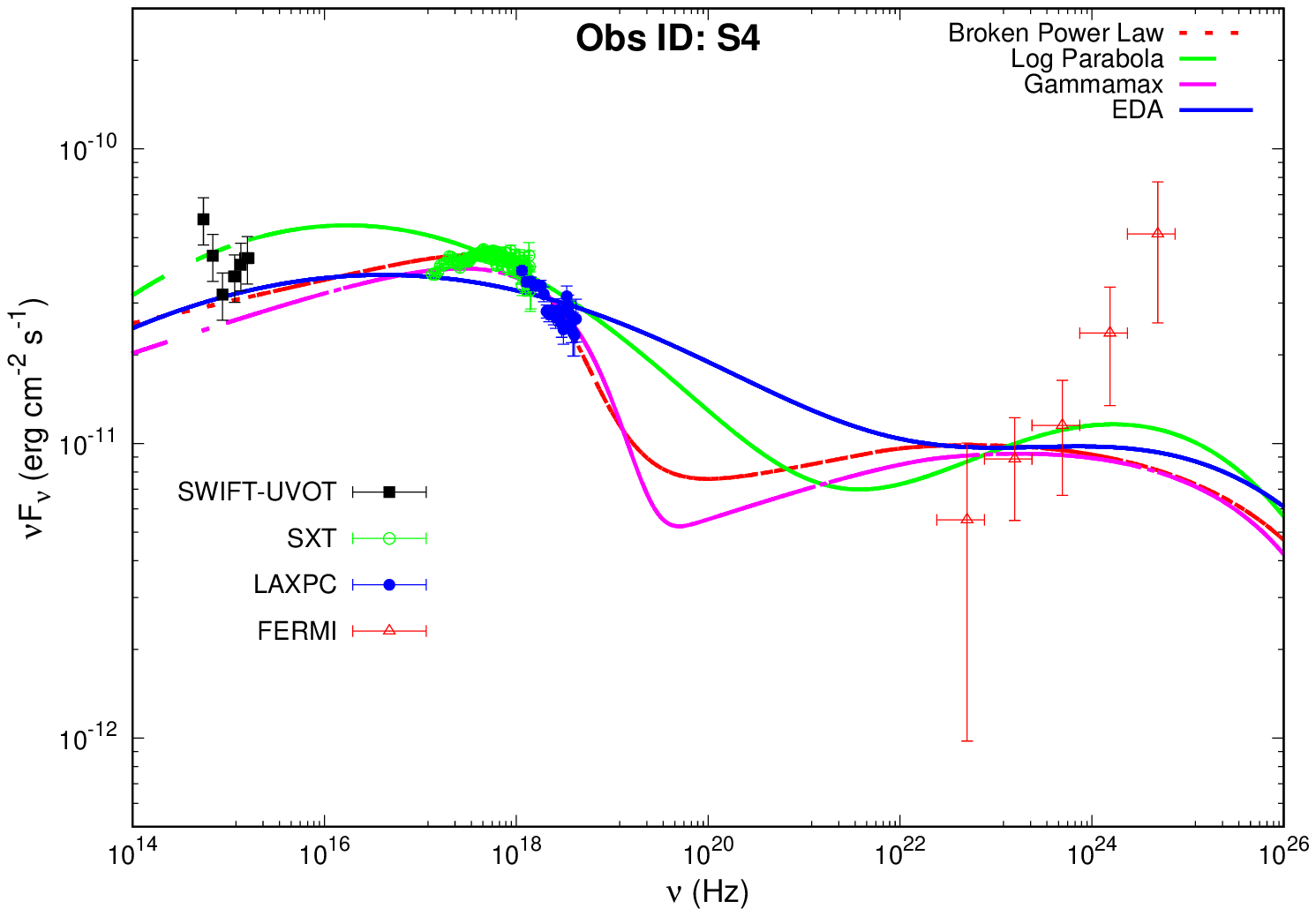}\\%
\includegraphics[scale=0.3,angle=270]{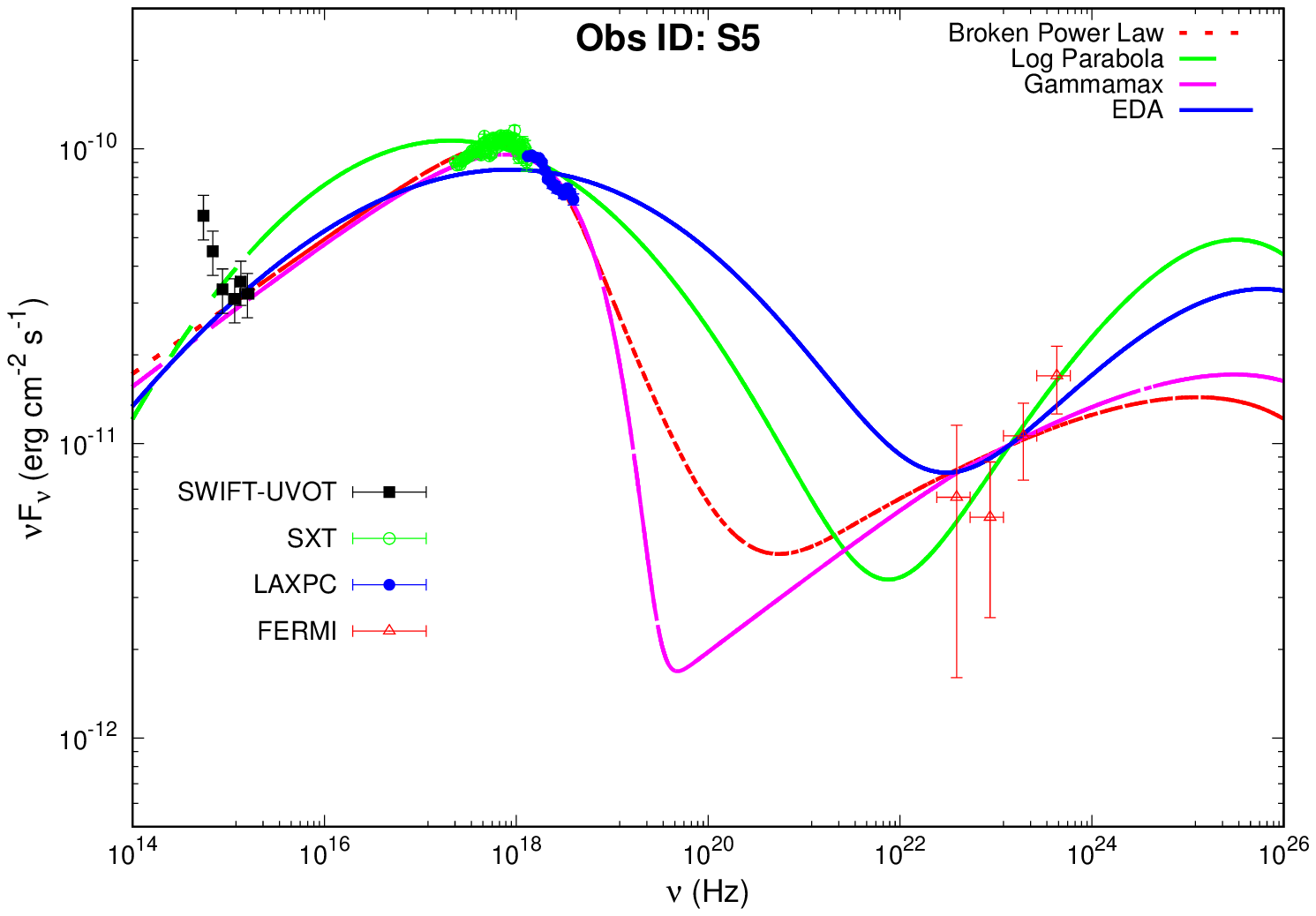}\\%

\vspace{12pt}

\caption{Broadband SED plots are generated for each observation using four distinct particle distribution models: LP, BPL, EDA, and $\xi_{max}$. The top-left plot corresponds to observation (S1), the top-right plot to observation (S2), the middle-left plot to observation (S3), the middle-right plot to observation (S4), and the bottom plot to observation (S5).}
\label{fig:sed_plots}

\end{figure*}

The limited amount of information from Optical/UV, X-ray, and $\gamma-$ray makes it difficult to constrain the full set of parameters.
 Consequently, we performed the broadband fitting by allowing the parameters $\Gamma$, B, $\alpha$, $\beta$, and Norm/$P_{jet}$ to vary when the particle distribution follows a logparabola model. In the case of a broken power-law model, the parameters $\Gamma$, B, particle indices before and after break energy ($\Gamma_{1}$ and $\Gamma_{2}$), and Norm/$P_{jet}$ were varied. For the EDA model, the parameters $\Gamma$, B, $\kappa$, $\psi$, and Norm/$P_{jet}$ were adjusted, and for the $\gamma$-max model, the parameters $\Gamma$, B, $p$, $\xi_{max}$, and Norm/$P_{jet}$ were varied. Other parameters for these models were fixed to typical values necessary for the observed broadband spectrum. For example, we have frozen the parameters R, $\Gamma$, and $\theta$ at $10^{17}$\,cm, 20, and 2 degree respectively in all the observations. 
These values are comparable with those used in previous studies conducted on the non-flaring emission of the source
  \citep{Abdo_2011, Anderhub_2009}.
Additionally, to achieve a reasonable reduced-$\chi^2$ in all the observations, we added a systematic error of 3\% to the X-ray and $\gamma$-ray data and 20\% to the Optical/UV data. We noted that the broadband SED fit with the four considered particle distributions yielded similar reduced-$\chi^2$ values for all observations. The reduced-$\chi^2$ values imply a degeneracy in the underlying particle distributions. The best-fit parameter values along with the reduced-$\chi^2$ for five observations are summarised in Table  \ref{table:4} and the broadband SED points along with model fits are shown in Figure ~\ref{fig:sed_plots}.

\begin{table*}
\centering
\scriptsize
\caption{Details of the best ﬁt parameters obtained by ﬁtting the broadband spectrum of Mrk\,501 with the one zone SED added as a local model in XSPEC. In the table, Columns 1: Observations,  2: Particle distribution models   3: Bulk Lorentz  factor,  4, 5, 6: represents the parameters corresponding to chosen particle distribution, 
7: Magnetic ﬁeld in units of Gauss, 8: Jet power in logarithmic scale in units of erg $sec^{-1}$,  9: Reduced-$\chi^2$. The subscript and superscript on the parameter values denote the lower and upper bound errors respectively, while -- indicates the lower or upper bound error is not constrained.}
 

\label{tab:5}
\begin{tabular}{c c c c c c c c c c c c}
\hline
\hline

Observation&Model&$\Gamma$&&&&B(G)&$P_{jet}$&$\chi^2_{red}$\\
(1) & (2) & (3) & (4) & (5) & (6) & (7) & (8)&(9)\\
\hline
&&&$\alpha$&$\beta$&&&\\
S1&LP &$20^{}_{}$&$3.54^{+0.05}_{-0.04}$&$0.48^{+0.08}_{-0.05}$&&$0.003^{+0.0008}_{-0.0007}$&$43.85^{+0.16}_{-0.25}$&0.96\\
\\
&BPL&&$\Gamma_{1}$&$E_{break}$&$\Gamma_{2}$&&&&\\ 
&&$20^{}_{}$&$2.90^{+0.05}_{-0.04}$&$1.30^{+0.2}_{-0.1}$&$5.08
^{+0.32}_{-0.30}$&$0.01^{+0.008}_{-0.005}$&$46.49^{+0.16}_{-0.23}$&0.73 \\
\\
&EDA&&$\kappa$&$\psi$&&&\\
&&$20^{}_{}$&$0.17^{+0.07}_{-0.01}$&$2.70^{+0.03}_{-0.03}$&&$0.005^{+0.001}_{-0.0007}$&$44.41^{+0.21}_{-0.28}$&0.86\\
\\
&$\xi_{max}$&& p & $\xi_{max}$&&&\\
&&$20^{}_{}$&$2.82^{+0.02}_{-0.01}$&$2.94^{+0.21}_{-0.20}$&&$0.01^{+0.002}_{-0.001}$ &$53.67^{-}_{-0.31}$&0.72\\ 
\\
\hline
&&&$\alpha$&$\beta$&&&\\
S2&LP&$20^{}_{}$&$3.62^{+0.02}_{-0.02}$&$0.45^{+0.005}_{-0.006}$&&$ 0.006^{+0.0005}_{-0.0004}$&$44.00^{+0.12}_{-0.19}$&1.34\\
\\
&BPL&&$\Gamma_{1}$&$E_{break}$&$\Gamma_{2}$&&&&\\
&&$20^{}_{}$&$3.00^{+0.03
}_{-0.05}$&$1.32^{+0.3}_{-0.1}$&$5.96
^{+2.54}_{-1.33}$&$0.01^{+0.002}_{-0.009}$&$46.71^{+0.11}_{-0.17}$&0.96 \\
\\
&EDA&&$\kappa$& $\psi$&&&\\
&&$20^{}_{}$&$0.14^{+0.003}_{-0.003}
$&$2.73^{+0.04}_{-0.04}$& --&$0.002^{+0.003}_{-0.0005}$&$44.81^{+0.27}_{-0.35}$&1.26\\
\\
&$\xi_{max}$&& p & $\xi_{max}$&&&\\
&&$20^{}_{}$&$2.88^{+0.01}_{-0.09}$&$2.58^{+0.28}_{-0.56}$&&$0.01^{+0.002}_{-0.001}$&$54.03^{-}_{-0.56}$&1.08 \\
\\
\hline
&&&$\alpha$&$\beta$&&&&&\\
S3&LP&$20^{}_{}$&$3.28^{+0.01}_{-0.02}$&$0.23^{+0.04}_{-0.04}$&&$ 0.005^{+0.002}_{-0.009}$&$45.10^{+0.16}_{-0.19}$&1.43\\
\\
&BPL&&$\Gamma_{1}$&$E_{break}$&$\Gamma_{2}$&&&&\\
&&$20^{}_{}$&$3.02^{+0.02}_{-0.01}$&$2.04^{+0.1}_{-0.2}$&$6.51
^{+0.87}_{-0.77}$&$0.01^{+0.002}_{-0.001}$&$46.90^{+0.14}_{-0.14}$&1.11\\
\\
&EDA&&$\kappa$& $\psi$&&&\\
&&$20^{}_{}$&$0.05^{+0.004}_{-0.005}
$&$2.32^{+0.03}_{-0.03}$&&$0.007^{+0.001}_{-0.009}$&$46.00^{-}_{-0.30}$& 1.33 \\
\\
&$\xi_{max}$&& p & $\xi_{max}$&&&\\
&&$20^{}_{}$&$2.93^{+0.01}_{-0.09}$&$4.42^{+0.16}_{-0.16}$&&$0.007^{+0.002}_{-0.0009}$&$54.37^{-}_{-0.40}$&1.18 \\

\\
\hline
&&&$\alpha$&$\beta$&&&&&\\
S4&LP &$20^{}_{}$&$3.36^{+0.03}_{-0.02}$&$0.43^{+0.01}_{-0.006}$&&$ 0.03^{+0.008}_{-0.008}$&$43.57^{+0.14}_{-0.19}$&1.06\\
\\
&BPL&&$\Gamma_{1}$&$E_{break}$&$\Gamma_{2}$&&&&\\
&&$20^{}_{}$&$2.97^{+0.02
}_{-0.01}$&$2.88^{+0.02}_{-0.01}$&$6.38
^{+1.20}_{-0.83}$&$0.01^{+0.001}_{-0.001}$&$46.50^{+0.10}_{-0.07}$&0.94\\
\\
&EDA&&$\kappa$& $\psi$&&&\\
&&$20^{}_{}$&$0.04^{+0.002}_{-0.004}
$&$2.10^{+0.02}_{-0.02}$&&$0.005^{+0.005}_{-0.001}$&$45.43^{+0.25}_{-0.48}$&1.36 \\
\\
&$\xi_{max}$&& p & $\xi_{max}$&&&\\

&&$20^{}_{}$&$2.78^{+0.01}_{-0.01}$&$6.30^{+0.61}_{-0.55}$&&$0.006^{+0.002}_{-0.001}$&$53.42^{-}_{-0.20}$&1.04\\
\\
\hline
&&&$\alpha$&$\beta$&&&&&\\
S5&LP&$20^{}_{}$&$2.73^{+0.02}_{-0.04}$&$0.31^{+0.02}_{-0.009}$&&$ 0.01^{+0.002}_{-0.001}$&$43.56^{+0.02}_{-0.05}$&0.75\\
\\
&BPL&&$\Gamma_{1}$&$E_{break}$&$\Gamma_{2}$&&&&\\
&&$20^{}_{}$&$2.54^{+0.01
}_{-0.02}$&$2.29^{+0.03}_{-0.02}$&$4.67
^{+1.1}_{-0.74}$&$0.03^{+0.001}_{-0.001}$&$45.67^{+0.12}_{-0.17}$&1.42\\
\\
&EDA&&$\kappa$& $\psi$&&&\\
&&$20^{}_{}$&$0.09^{+0.002}_{-0.004}
$&$1.94^{+0.02}_{-0.02}$&&$0.005^{+0.001}_{+0.004}$&$44.82^{-}_{-0.22}$&1.20 \\
\\
&$\xi_{max}$&& p & $\xi_{max}$&&&\\
&&$20$&$2.50^{+0.02}_{-0.01}$&$5.07^{+0.51}_{-0.44}$&&$0.009 ^{+0.001}_{-0.002}$&$52.60^{+0.42}_{-0.42}$&1.45\\
\\
\hline 
\hline
\label{table:4}
\end{tabular}
\end{table*}
\subsection{Correlation analysis}
\begin{figure*}
\centering

\includegraphics[scale=0.5]{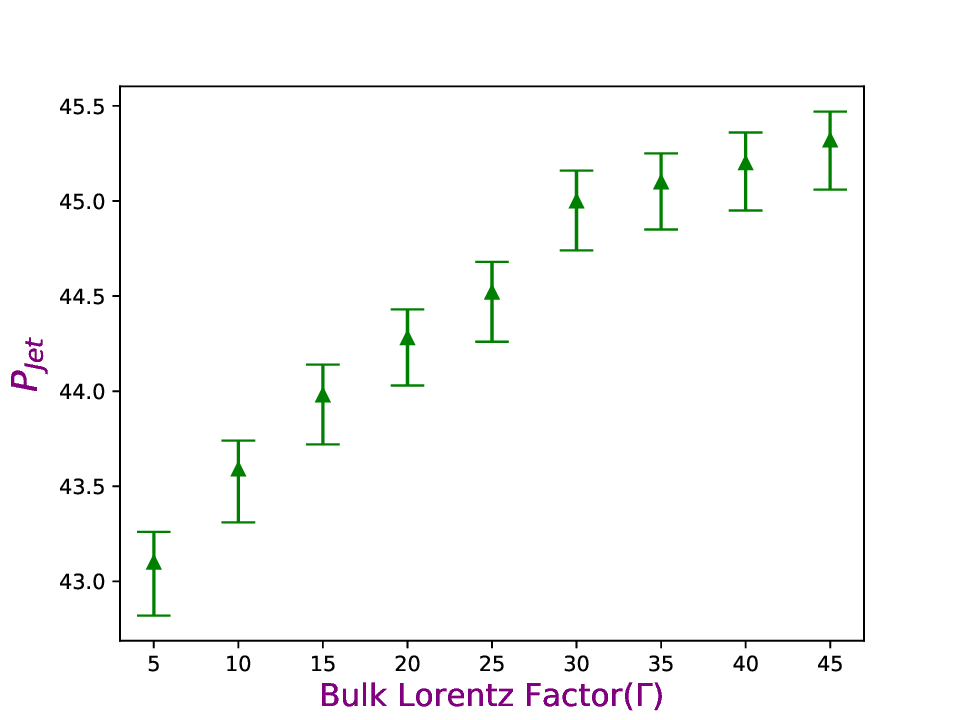}%
\includegraphics[scale=0.5]{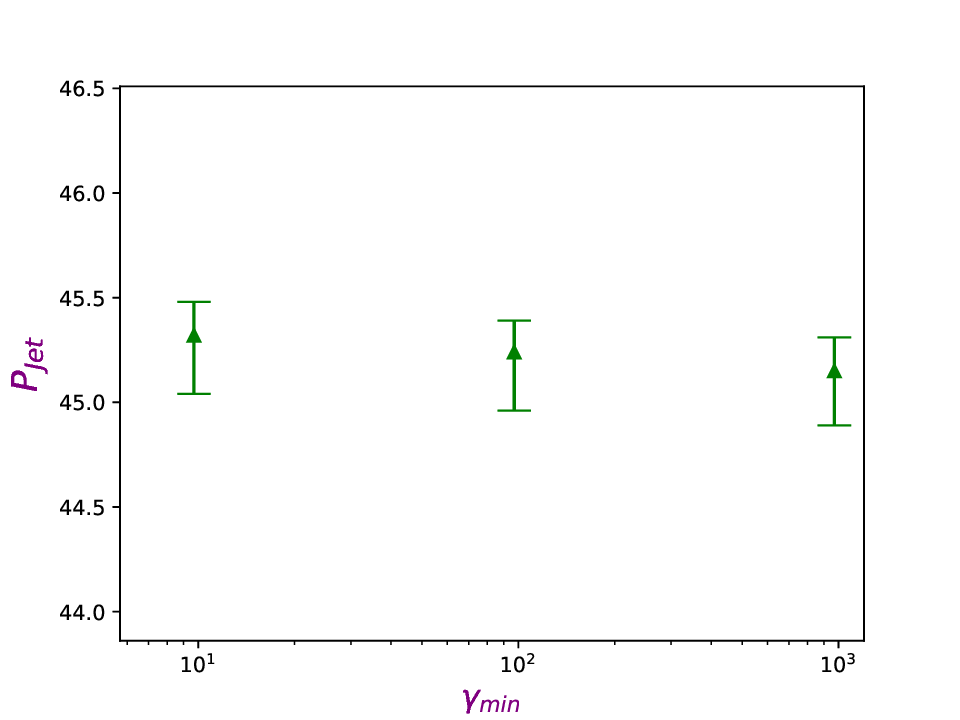}\\
\includegraphics[scale=0.5]{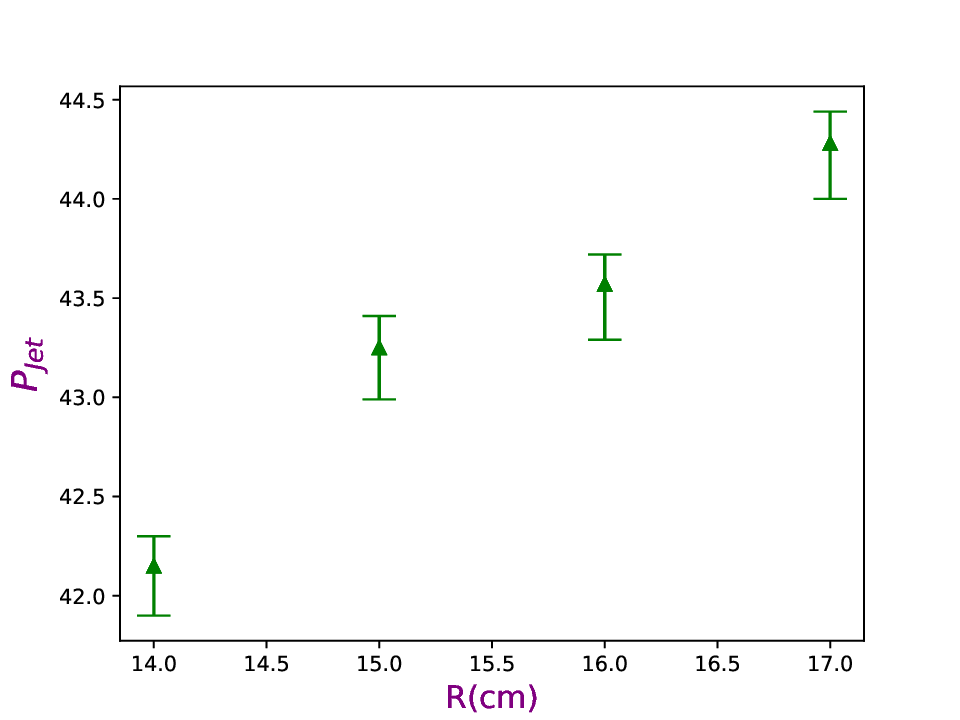}
\includegraphics[scale=0.5]{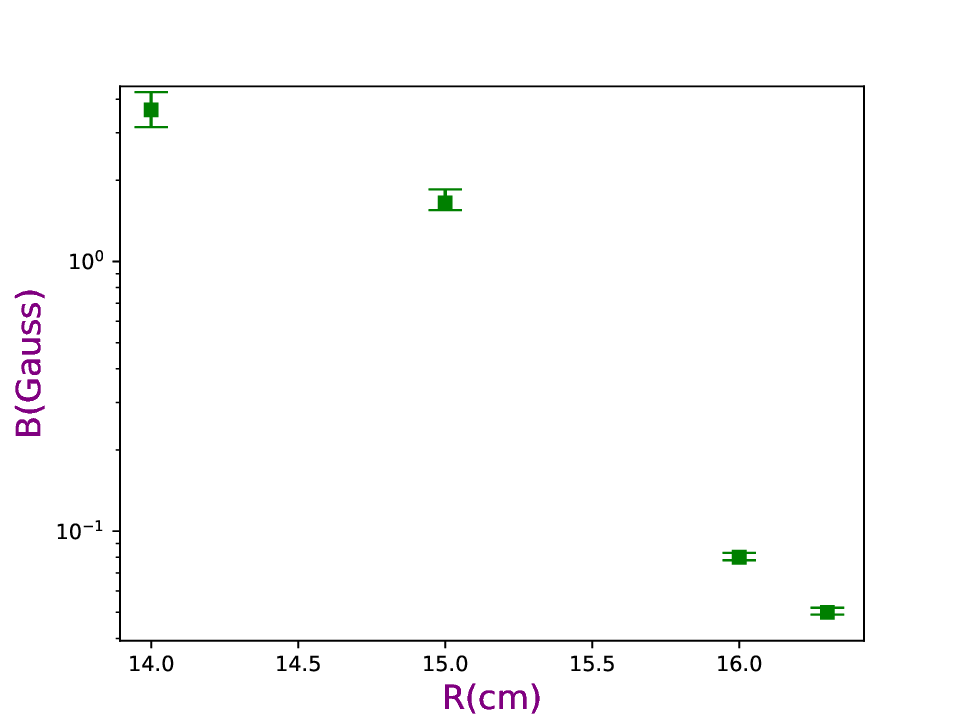}\\
\caption{Jet power variation with the bulk Lorentz factor, minimum particle energy and emission region size using the LP model for the particle distribution. 
Top left panel is a variation of $\rm P_{jet}$ with $\Gamma$, top right panel is a plot of jet power with $\gamma_{min}$, the bottom left panel is a plot of $\rm P_{jet}$ with R. Bottom right panel represents the variation of R with a B.}
\label{fig:cor_log}
\end{figure*}

\begin{figure*}
\centering
\includegraphics[scale=0.5]{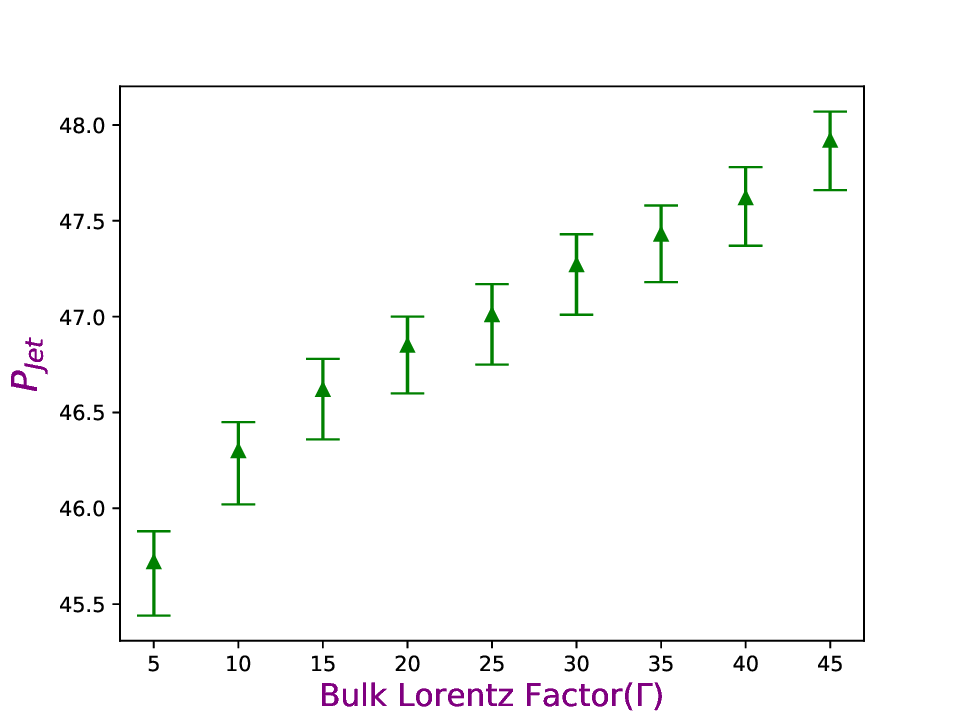}
\includegraphics[scale=0.5]{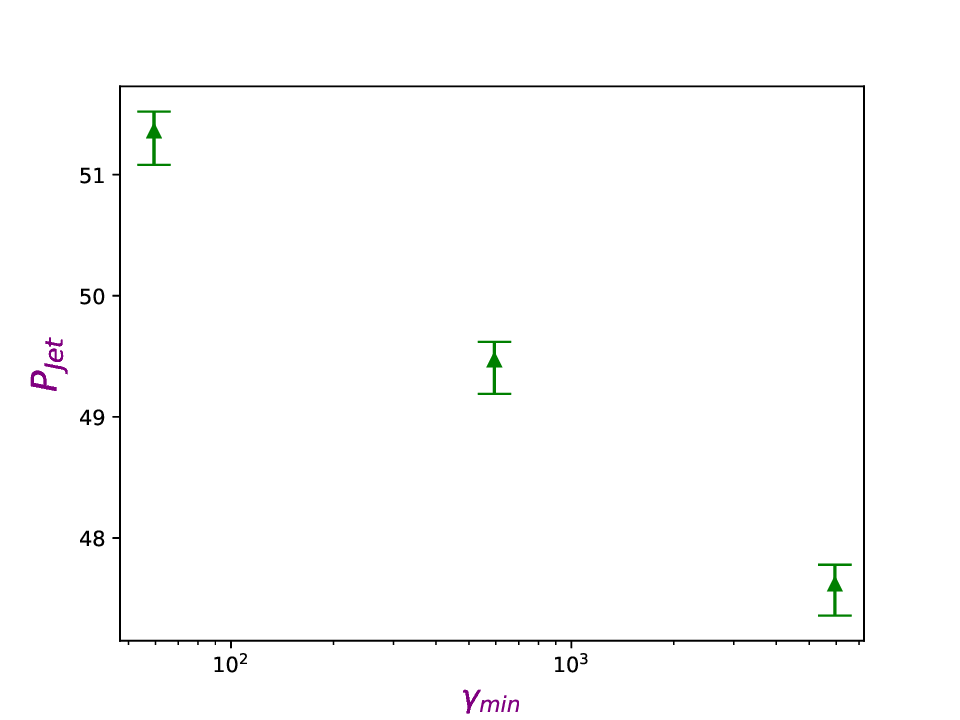}\\
\includegraphics[scale=0.5]{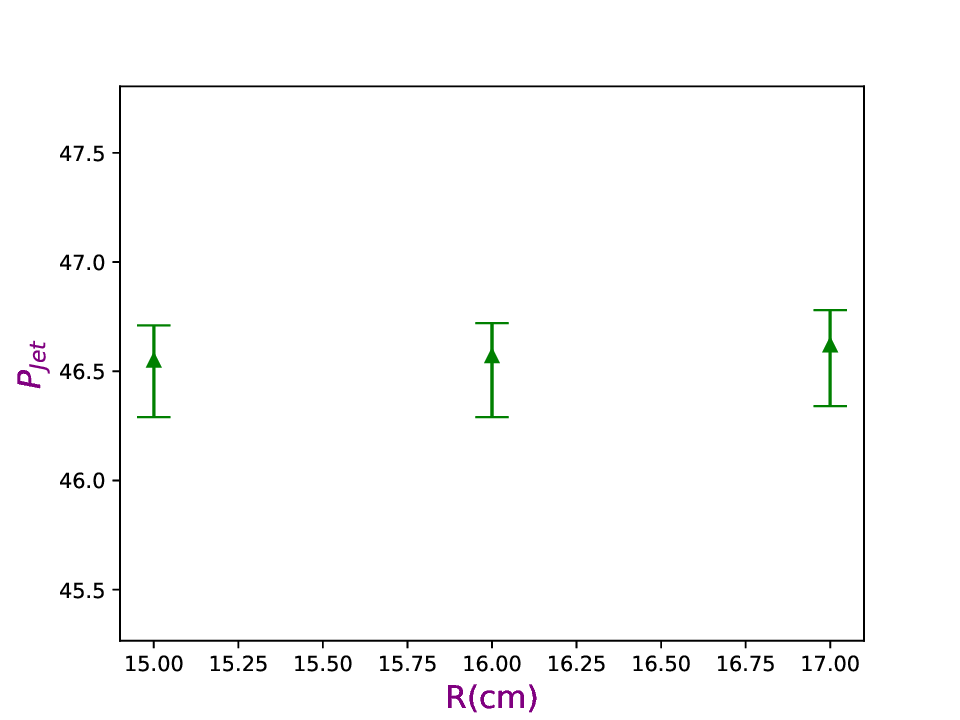}
\includegraphics[scale=0.5]{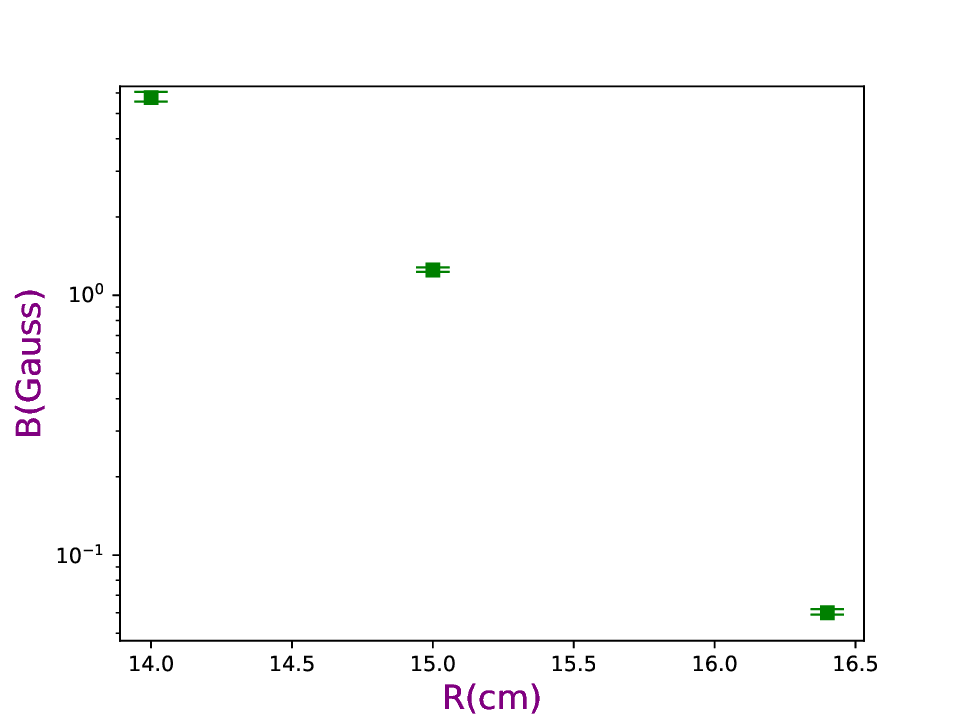}\\
\caption{ Jet power variation with the bulk Lorentz factor, minimum particle energy and emission region size using the BPL model for the particle distribution. The top left panel is a variation of $\rm P_{jet}$ with $\Gamma$, top right panel is a plot of jet power with $\gamma_{min}$, the bottom left panel is a plot of $\rm P_{jet}$ with R.   Bottom right panel represents the variation of R with a B.}
\label{fig:cor_bkn}

\end{figure*}

 The four models for the particle distribution yielded similar reduced -$\chi^2$ values, indicating a degeneracy in the underlying particle distributions. However, the best-fit parameters and properties acquired from these models can be used to remove the degeneracy and rule out some of the models. Specifically, the jet power estimated from the $\xi_{max}$ model shows significantly higher values in all the observations compared to those obtained using other forms of particle distribution. Consequently, based on the criterion of minimum jet power, we can exclude the $\xi_{max}$ model as a plausible form of particle distribution. Additionally,  the LP model being an approximation of the EDA model \citep{Hota:2021csa}, implies that the underlying mechanism responsible for the particle distribution is the same in both models. Therefore, it is equivalent to consider either of the models. Due to the simpler form of the LP, we opted for this model and the BPL model for subsequent analysis. To further constrain the particle distribution and accurately estimate the $\rm P_{jet}$ in different observations of Mrk\,501, we conducted a comprehensive correlation analysis of the $\rm P_{jet}$ with the $\Gamma$, $\gamma_{min}$ and R (see Figure ~\ref{fig:cor_log} and~\ref{fig:cor_bkn}). In the correlation plots , we calculated $\rm P_{jet}$  for different values of $\Gamma$, $\gamma_{min}$ and R.  We also explored the correlation of R with the magnetic field (right bottom panel of Figures ~\ref{fig:cor_log} and~\ref{fig:cor_bkn}), here magnetic filed is estimated for different values of R. 
  Notably, we observed that the $\rm P_{jet}$ consistently increased with a rise in  $\Gamma$ for both models. Next, we fixed $\Gamma\sim 20$ and calculated the $\rm P_{jet}$ for different values of minimum particle energy. In the LP model, we found that the $\rm P_{jet}$ is independent of $\gamma_{min}$. While in the BPL model, the $\rm P_{jet}$ exhibited a decrease trend with an increase in $\gamma_{min}$. We further investigated the impact of varying the size (R) on the $\rm P_{jet}$. In the LP model, $\rm P_{jet}$ increased with a rise in R, whereas in the BPL model, $\rm P_{jet}$ remained constant for different R values. Additionally, our analysis revealed that, in both models, the size of the emission region required decreased with an increase in the magnetic field.

\section{Summary and Discussion}
\label{sum:dis}
The BL\,Lac object, Mrk\,501, has been observed by the \emph{Fermi}-LAT instrument since its launch. For the first time, these observations together with the simultaneous observations of \emph{Astro}Sat and \emph{Swift}-UVOT have enabled a comprehensive temporal and spectral study of the source. Our analysis of the 3-day binned $\gamma-$ray light curve revealed that the emission during this period MJD 57615 -- 59666 corresponded to a quiescent state of the source. These results made Mrk\,501 an important blazar source to understand the underlying physics. Since in most blazars, the multifrequency data is available in high flux states, and very sparse data is available in low flux states. This is primarily because multifrequency campaigns are triggered by observing enhanced flux levels in some energy bands. Hence much of our understanding of the blazars is biased towards “high-activity” states, where perhaps distinct physical processes play a dominant role. Hence, it is important to study the source in the low flux states. In the case of Mrk\,501, the long-duration observations are available in the low flux state from multiple Telescopes. In the X-ray range, \emph{Astro}Sat has archived five observations during the period  MJD 57615 -- 59666. These observations suggest that Mrk\,501 was predominantly in a low flux state, though the flux (counts/second) exhibited a slight increase during the period MJD 58910.3 -- 59665.8. A recent study conducted by \citep{abe2023multi} has also indicated a historically low flux for Mrk\,501 from mid-2017 to mid-2019. The long-lasting stable emission can potentially be explained by a standing shock scenario proposed by \citep{marscher2008} and further explored in \citep{marscher2014}. 
Our analysis of the temporal behavior of Mrk\,501 using SXT and LAXPC light curves showed notable variability during the source's low flux state. A comparison of variability across different energy bands indicated that the maximum variability occurred in the X-ray band as compared to the optical and $\gamma$-ray band. These findings are consistent with previous studies conducted during both low and high flux states. \citet{2019Galax...7...62S} analyzed data from different instruments, including FACT, Swift
XRT/BAT, and Fermi telescopes, over the period from January 2013 to June 2018. The study observed that variability is significantly larger in X-rays compared to $\gamma$-rays in both Mrk\,501 and Mrk\,421. Additionally, in the VHE regime, the variability was found to be more pronounced
in Mrk\,501 than in Mrk 421. During the 2008 campaign on Mrk\,501, it was observed that variability was greater in X-rays and VHE compared to optical
wavelengths \citep{2015A&A...573A..50A}. More recent studies by \citet{abe2023multi, Kapanadze_2023} examined the source during a low flux state. Similarly, they found that X-ray variability is larger than that observed in $\gamma$-ray and optical data. These findings are consistent with our studies and further validate our conclusions on the significant variability observed in X-rays across different flux states.
 In Mrk\,501, the amplitude of variability in distinct energy bands  corresponds to the shape of the broadband SED. Given that the X-ray spectrum in the broadband SED lies around the peak, it is therefore mostly associated with high-energy electrons, while the $\gamma$-ray and optical spectra, situated before the break energy, are due to low-energy electrons. The faster cooling of high-energy electrons emitting X-rays contributes to a larger variability amplitude in the X-ray band, whereas the slower cooling of low-energy electrons results in a smaller variability amplitude in the Optical/UV and $\gamma$-ray bands.

We know that the power-law (PL) spectral shape is the characteristic feature of non-thermal emission \citep{landau1986}.  \cite{Anderhub_2009}  analyzed this source in a low flux state and found that the X-ray spectral shape derived from the Suzaku data from 0.6 keV to 40 keV was well described by a broken power law.  However, recent observations of some blazars with sensitive Telescopes show that the X-ray spectra significantly deviate from a power-law shape and instead, they show a mild curvature in their spectra \citep{massaro2008,Furniss_2015, Hota:2021csa}.  Our study confirms that both log-parabola and broken power law models provide a good fit to the data.  Using the \emph{Astro}Sat observation, we conducted a comprehensive X-ray spectral analysis of Mrk\,501. The wide-band X-ray spectrum from LAXPC and SXT instruments was jointly fitted using PL, BPL, and LP models. We noted that the PL model results in high reduced-$\chi^2$ values compared to the BPL and LP models. However, both BPL and LP models produced similar fits with comparable reduced-$\chi^2$ values across all observations.  
The hard spectrum of the source corresponds to the brightest observation considered in our work, indicating a harder brighter feature (see Figure ~\ref{fig:meancount}).  
The source clearly shows a “harder when brighter” trend even during the low flux states. This feature is typically observed during the flaring states of blazars. 
 In the case of LP, the spectral indices defined at the pivot energy, $\alpha$ ranged from 1.66 -- 2.20 and the curvature parameter $\beta$ ranged from 0.18 -- 0.36. These values indicate that the energy spectrum of the source exhibits moderate curvature in the low flux states. Notably, the parameters of the LP model are consistent with those of the BPL model, with the $\alpha$  being hardest during the bright observation. Additionally, the observation with the hardest spectrum exhibits a maximum curvature value of 0.36.  The observed curvature cannot be simply attributed to the radiative cooling of high-energy electrons, such as synchrotron and inverse Compton processes. Instead, it is more likely linked to the particle distribution in the emission region resulting from stochastic acceleration, particle diffusion, or electron escape mechanisms. The standard approach to explain the curvature involves the combined effects of particle acceleration and radiative losses. For instance, \citep{Massaro2004} demonstrated that a curved spectrum can result from an energy-dependent acceleration process, where the acceleration probability decreases with electron energy. Alternatively, an energy-dependent electron escape rate from the acceleration region can also lead to the curvature of the emitted spectrum \citep{10.1093/mnrasl/sly086, Goswami_2018}.
In our work, we too assumed that the intrinsically curved particle distributions are due to an energy dependence of the acceleration time-scales. However, a more comprehensive examination of the micro-physics involved in these processes is necessary to confirm the assumed dependence.

{\bf We modeled the broadband SED of Mrk\,501 by using the convolved one-zone leptonic model involving the synchrotron and SSC processes.  A key advantage of our model is its integration with XSPEC, allowing for rigorous statistical fitting and providing well-defined confidence intervals on the model parameters. Additionally, our code is flexible, as it can accommodate any user-defined particle distribution, including those derived from solving the Fokker-Planck equation, and easily convolve it with the broadband emission model. This provides a significant edge over other available models, where incorporating custom particle distributions can be more complex. Another important advantage is that XSPEC offers an accessible platform for constructing and customizing models, simplifying the process compared to more complex frameworks like JetSet \citep{2011ApJ...739...66T}. Additionally, while the other codes like LeHaMoC \citep{2024A&A...683A.225S}  and SOPRANO \citep{2022MNRAS.509.2102G}  focus on modeling, they often lack the capability to provide statistical uncertainties on the parameters, whereas our method fully utilizes XSPEC’s statistical fitting capabilities to ensure a more comprehensive analysis. 
 
The Fermi-LAT $\gamma$-ray data from Mrk\,501 were analysed using Fermitools v2.2.0, following the standard procedures outlined in the Fermi-LAT documentation \citep{2017ICRC...35..824W}. Similarly, we processed \emph{Swift}-UVOT into scientific products using the HEASoft packages namely \textit{UVOTSOURCE} and \textit{UVOTIMSUM}. After extracting the flux points and energies, we converted them into XSPEC-compatible PHA format using the ftflx2xsp command. This allowed us to load the PHA files into XSPEC for the broadband spectral analysis.}
 We considered different forms of particle distribution like BPL, LP, EDA, and  $\xi_{max}$  in our analysis for the broadband SED modelling. Initially, we freezed  $\Gamma$, R, and $\theta$ at 20, $10^{17}$\,cm and 2 degree, respectively, and varied other parameters.  We noted that the four models for the particle distribution yielded similar reduced-$\chi^2$ values and the $\rm P_{jet}$ varied among the models. The $\rm P_{jet}$ values in units of erg/s, ranged from 43.56 - 45.10 (LP Model), 45.67 - 46.90 (BPL Model), 44.41 - 46.00 (EDA Model), and 52.60 - 54.03 ( $\xi_{max}$). Notably, $\rm P_{jet}$ from the $\xi_{max}$ model was consistently higher across observations, whereas it was comparatively lower for the LP and EDA models.  \citet{10.1093/mnras/stae706} also studied the $\rm P_{jet}$ of Mrk\,501 during the low flux state, and similarly observed lower $\rm \, P_{jet}$ values in the LP and EDA models, with their reported ranges being 42.18 - 44.03 (LP Model), 46.6 - 48.01 (BPL Model), 42.65 - 45.06 (EDA Model), and 46.5 - 48.19 ($\xi_{max}$ Model), which differ somewhat from our findings.
 Based on the criterion of minimum jet power, we excluded the $\xi_{max}$ model as a plausible form of particle distribution responsible for the emission. Since the LP and EDA models exhibited the lowest $\rm P_{jet}$, hence we chose the simpler LP model for the correlation analysis alongside the BPL model.
Comparing our LP model parameters with those from Bora et al. (2024), we found that our $\beta$ values (0.23 - 0.48) had a narrow range compared to the $\beta$ values (0.24 - 0.79) obtained in \citet{10.1093/mnras/stae706} work. However, the  $\alpha$ values (2.73 - 3.62) obtained in our study were consistent with the $\alpha$ values  (2.77 - 3.81) reported by  \citet{10.1093/mnras/stae706}.
In our work, the magnetic field values obtained using the LP model range from 0.003 to 0.03 G, and remain approximately 0.01 G in all flux states in the case of the BPL model.
\\
It is important to note that TeV observations are crucial for constraining jet properties in HBLs through SED modeling. However, no simultaneous data are available for Mrk 501 during our analysis period, with the closest TeV observation separated by three months. HBLs, being low-luminosity sources with featureless spectra, are believed to scatter only synchrotron photons to high-energy $\gamma$-rays without significant contributions from external photons \citep{2008MNRAS.387.1669G, 2016ApJ...831..142M}. In this context, a leptonic model including synchrotron and SSC components is adequate for modeling broadband emission. Importantly, we noted that the SED parameters obtained in our work are consistent with previous results that incorporated VHE emission in the SED modeling. For example, \cite{Abdo_2011} studied this source in a low flux state with VHE emission included and found broadband SED parameters, such as a magnetic field of 0.03 G and an emission region size of  $10^{17}$
 cm.  In our work, we have used similar values for the broadband SED modeling.  Interestingly,  low values of the magnetic field are also reported in previous works \citep{Kataoka_1999, Anderhub_2009, abe2023multi, Abe:2024eba}. Additionally, \citet{Furniss_2015} examined Mrk\,501 in both low and high states and found that the SSC model could reproduce the observed broadband states through a decrease in magnetic field strength coupled with an increase in the luminosity and hardness of the relativistic leptons. This study also noted that the high flux state has a lower magnetic field (0.03 G) than the low flux state magnetic field (0.05 G). These studies highlight that  our results remain comparable with previous studies which include VHE emission. Recently, \citet{abe2023multi} used multiwavelength observations to complement IXPE pointings of Mrk 501, demonstrating that changes between different states can be explained by variations in the magnetization of the emission region. Similarly, \citet{2020A&A...637A..86M} noted that during the maximum X-ray activity of Mrk\,501 in July 2014, the VHE $\gamma$-rays showed an elevated emission. The authors demonstrated flux variations was mainly due to changes in the break energy. These results indicate that variations in the flux of the source are mostly related to the change in the break energy and magnetic field of the emission region.

We studied the dependence of  $\rm P_{jet}$ with  $\Gamma$, $\gamma_{min}$ and R.
 Our study revealed a consistent increase in $\rm P_{jet}$ with a rise in $\Gamma$ for both the models. When $\Gamma$ was fixed at $\sim 20$, we found that  $\rm P_{jet}$ is independent of $\gamma_{min}$ in the case of LP model. However, in the BPL model, the $\rm P_{jet}$ decreased as $\gamma_{min}$ increased. Additionally, we explored the impact of  R on the $\rm P_{jet}$, $\rm P_{jet}$ increased with an increase in the values of R.
While in the BPL model, $\rm P_{jet}$ remained constant for different values of size of the emission region. 
 Our analysis also revealed that, in both models, the required size (R) decreased as B increased.
 
 The comparatively low jet power acquired for the LP/EDA electron distribution models,  $\sim 10^{44}$ erg s$^{-1}$ makes it about 10 percent of the Eddington luminosity for a $10^7$ M$_\odot$ blackhole, $L_\text{Edd} \sim 10^{45}$ erg s$^{-1}$. This implication is significant, suggesting that such a jet could be directly fueled by the accretion process. This, in turn, offers a potential framework for studying the connections between jets and accretion discs \citep{10.1046/j.1365-8711.1999.02657.x, Ghisellini_2014, 10.1093/mnras/stw1730}.  Interestingly, \citet{Deng:2021twz}  also showed that for Mrk 501, the power resulting from the rotating black hole is systematically higher than the jet power, while the jet power is comparable to the power originating from the accretion disk.
Moreover, previous studies have demonstrated a correlation between jet power and accretion disk luminosity \citep{10.1093/mnras/286.2.415, 1999PPMtO..18..243C, 2014Natur.515..376G, chen2015black, Xiao_2022}, further supporting the connection between jet power and the accretion disk. However, it is commonly observed that in BL Lac objects, especially the HBL class, the jet power often exceeds the disk luminosity \citep{2014Natur.515..376G, 2024A&A...682A.134G}. For example, a recent study by \citet{2024A&A...682A.134G} examined the behavior of various HBL sources by modeling their SEDs using three approaches and found that jet power exceeds disk luminosity for HBLs in all three methods. In these sources, the strong broad emission lines, which would typically indicate accretion disk luminosity, are generally absent. As a result, the jet power in these sources is much greater than their accretion disk luminosity \citep{10.1093/mnras/stt2246}. These findings suggest that the jet power could originate from the spin and mass of the central black hole. The mechanism proposed by \citet{1977MNRAS.179..433B} explains this phenomenon through the torque exerted on the rotating black hole by the magnetic field amplified by the accreting material.
Thus, despite our findings, the question of what powers the jet in Mrk\,501 remains unclear and warrants further investigation.\\

The lower jet power also implies a small energy reservoir in the jet, making it susceptible to depletion from shocks and radiative losses within a shorter time frame. Consequently, there is a motivation for conducting a detailed investigation into the temporal evolution of a sample of blazars whose spectrum is intrinsically characterized by the curved particle distributions and hence obtain constraints on the jet power.


\newcommand{\aap}{Astronomy \& Astrophysics}
\newcommand{\apj}{The Astrophysical Journal}
\newcommand{\ssr}{Space Science Reviews}
\newcommand{\mnras}{Monthly Notices of the Royal Astronomical Society}
\newcommand{\apjl}{The Astrophysical Journal Letters}
\newcommand{\pasp}{Publications of the Astronomical Society of the Pacific}




\section{Acknowledgements}
 We sincerely thank the anonymous referees for their valuable feedback and suggestions, which have significantly improved the quality of our manuscrip. JT, ZS and NI are thankful to the Indian Space Research Organisation, Department of Space, Government of India (ISRO-RESPOND) for the financial support. ZS is supported by the Department of Science and Technology, Govt. of India, under the INSPIRE Faculty grant (DST/INSPIRE/04/2020/002319). We express our gratitude to the Inter-University Centre for Astronomy and Astrophysics (IUCAA) in Pune, India, for the support and facilities provided.

\bibliographystyle{elsarticle-harv} 
\bibliography{example}






\end{document}